\documentclass[manuscript]{aastex63}
\usepackage{booktabs} 
\usepackage{pgfplotstable}
\usepackage{csvsimple}
\usepackage{multirow}

\makeatletter

\newcommand{\Rmnum}[1]{\expandafter\@slowromancap\romannumeral #1@}
\makeatother
\graphicspath{{./}{fig/}}
\usepackage{threeparttable}
\usepackage{dblfnote}
\pgfplotsset{compat=1.17}

\shorttitle{Very High-$z$ SNe \Rmnum{1}a and Cosmological Parameters}

\begin{document}
\title{Constraints on Cosmological Parameters with a Sample of Type Ia Supernovae from JWST}

\correspondingauthor{Lifan Wang}
\email{lifan@tamu.edu}

\author{Jia Lu}
\affil{Purple Mountain Observatory, Chinese Academy of Science, Nanjing, Jiangsu, 210033, China}
\affil{School of Astronomy and Space Science, University of Science and Technology of China, Hefei, Anhui, 230026, China}
\author{Lifan Wang}
\affil{Mitchell Institute for Fundamental Physics \& Astronomy, Texas A\&M University, Department of Physics and Astronomy, 4242 TAMU, College Station, TX 77843, USA}
\author{Xingzhuo Chen}
\affil{Mitchell Institute for Fundamental Physics \& Astronomy, Texas A\&M University, Department of Physics and Astronomy, 4242 TAMU, College Station, TX 77843, USA}
\author{David Rubin}
\affil{Department of Physics and Astronomy, University of Hawai`i at M\=anoa, Honolulu, HI 96822, USA}
\author{Saul Perlmutter}
\affil{E.O. Lawrence Berkeley National Laboratory, 1 Cyclotron Road, Berkeley, CA 94720, USA}
\affil{Department of Physics, University of California Berkeley, Berkeley, CA 94720, USA}
\author{Dietrich Baade}
\affil{European Organisation for Astronomical Research in the Southern Hemisphere (ESO), Karl-Schwarzschild-Str.\ 2, 85748 Garching b.\ M\"unchen, Germany}
\author{Jeremy Mould}
\affil{Centre for Astrophysics \& Supercomputing, Swinburne University, P.O. Box 218, Hawthorn, Vic 3122, Australia}
\author{Jozsef Vinko}
\affiliation{Department of Astronomy, University of Texas at Austin, 2515 Speedway, Stop C1400
Austin, Texas 78712-1205, USA}
\affil{Konkoly Observatory, CSFK, Konkoly Thege M. ut 15-17, Budapest, 1121, Hungary}
\affiliation{ELTE E\"otv\"os Lor\'and University, Institute of Physics, P\'azm\'any P\'eter s\'et\'any 1/A, Budapest, 1117 Hungary}
\affiliation{Department of Optics \& Quantum Electronics, University of Szeged, D\'om t\'er 9, Szeged, 6720, Hungary}
\author{Enik\H o Reg\H os}
\affiliation{Konkoly Observatory, CSFK, Konkoly Thege M. ut 15-17, Budapest, 1121, Hungary}
\affiliation{Magdalen College, Oxford, OX1 4AU, UK}
\author{Anton M. Koekemoer}
\affiliation{Space Telescope Science Institute (STScI), 3700 San Martin Drive Baltimore, MD 21218, USA}



\begin{abstract}
 
We investigate the potential of using a sample of very high-redshift ($2\lesssim z \lesssim6$) (VHZ) Type Ia supernovae (SNe~Ia) attainable by the James Webb Space Telescope (JWST) on constraining cosmological parameters.
At such high redshifts, the age of the universe is young enough that the VHZ SNIa sample comprises the very first SNe~Ia of the universe, with progenitors among the very first generation of low-mass stars that the universe has made. 
We show that the VHZ SNe~Ia can be used to disentangle systematic effects due to the luminosity distance evolution with redshifts intrinsic to SNIa standardization. Assuming that the systematic evolution can be described by a linear or logarithmic formula, we found that the coefficients of this dependence can be determined accurately and decoupled from cosmological models. Systematic evolution as large as 0.15 mag and 0.45 mag out to $z=5$ can be robustly separated from popular cosmological models for linear and logarithmic evolution, respectively. The VHZ SNe~Ia will lay the foundation for quantifying the systematic redshift evolution of SNIa luminosity distance scales. When combined with SNIa surveys at comparatively lower redshifts, the VHZ SNe~Ia allow for a precise measurement of the history of the expansion of the universe from $z\sim 0$ to the epoch approaching reionization. 

\end{abstract}


\section{Introduction} \label{sec:intro}

The accelerating expansion of the universe revealed by the observations of Type Ia supernovae (SNe~Ia) has been one of the most exciting discoveries in astronomy  \citep{Perlmutter1999ApJ...517..565P,1998AJ....116.1009R}.
Its primary results have been consistently strengthened by other observations such as the cosmic microwave background \citep[CMB; ][]{2001PhRvL..86.3475J}, baryon acoustic oscillations \citep[BAO;][]{EisensteinBAO2005ApJ...633..560E,2014MNRAS.441...24A}, and weak gravitational lensing \citep[WL; ][]{2018PhRvD..98d3526A}. The physics behind the acceleration, however, remains poorly understood. 
The observational data from SNe~Ia, BAO, and CMB \citep{Planck182020A&A...641A...6P} are remarkably consistent with the Flat $\Lambda$-cold-dark-matter ($\Lambda$CDM) model, which is now widely considered to be the standard cosmological model. In this model, the accelerating expansion of the universe can be explained by a cosmological constant that describes the vacuum energy with an equation of state of $w=-1$.
However, there are still unsolved problems in the standard $\Lambda$CDM model: the scale problem - the magnitude of the dark energy density measured by observations is much smaller than that predicted by quantum fluctuation theory by a factor of order $10^{56}$ \citep{1989RvMP...61....1w, 2005PhRvL..95n1301C}, the coincidence problem - why the magnitude of dark energy density shares the same order with the matter density today \citep{2003RvMP...75..559P}, and whether the vacuum energy or dark energy equation of state are constant or time-dependent \citep{2011ARNPS..61..251G}. 

To resolve these problems, many alternative theoretical models have been proposed. These models can be divided into dark energy models if a new form of matter has been considered - mathematically, the right side of the Einstein equations is modified, and modified gravity models, if a new form of force has been added - mathematically the left side of the Einstein equations, is modified \citep{2016ARNPS..66...95J}.

Various observations are needed to determine the best model describing the universe. The widely used observations include SNe~Ia, the WL, BAO, CMB, and clusters of galaxies (CL). Other probes like gravitational waves (GW) and long-duration gamma-ray bursts (LGRB) can offer supporting constraints \citep{2008ARA&A..46..385F}.
SNe~Ia are the most accurate cosmic distance candles to probe cosmic acceleration. With the ever-increasing number of well-observed SNe~Ia, the statistical errors associated with SNe~Ia have been reduced largely, and the main uncertainties on cosmological parameter estimation are dominated by systematic errors now \citep[e.g.][]{2019astro2020T.270S}. 
Most currently available SNIa data are from redshifts below $2$. Extending this redshift range to well beyond $2$ has been impossible but the situation will change with the launch of the James Webb Space Telescope (JWST) which will enable discoveries of SNe~Ia at redshifts approaching the epoch of reionization \citep{Wang2017arXiv171007005W,Regos2019ApJ...874..158R}. The SNe~Ia at such high redshifts are likely to be from systematically young and metal-poor progenitor systems, which allow for some critical systematic effects of SNIa luminosity distance standardization to be studied. The SNe~Ia at redshifts beyond $2$ probe the universe well before dark energy dominance \citep{NaoSuzuki2012,Scolnic2018ApJ...859..101S} and can place precise constraints on the matter density of the universe, but the most popular cosmological models are less sensitive to the physics of dark energy. While this insensitivity makes these very high redshifts (VHZ) SNe~Ia only indirectly related to dark energy, it actually simplifies the background cosmological models for the studies of the intrinsic properties of SNe~Ia and allows the physics intrinsic to SNe~Ia to be decoupled from dark energy driven cosmological models. \cite{RiessLivio2006ApJ...648..884R} show that at redshift from 1.5 to 3.0 the systematic evolution of SNe Ia can be the strongest if the delay time of SN Ia explosions from the formation of their progenitors is around 2-3 Gyrs. However, a major portion of SNe Ia may explode with a delay time as short as 200 million years \citep[e.g.,][]{Castrillo2021MNRAS.501.3122C,ChenXingzhuoMuse2021arXiv210106242C,Wiseman:10.1093/mnras/stab1943}. It is thus important to observe SNe at even higher redshifts. 

Transient surveys in the near future will produce a much larger sample of well-observed SNe~Ia. The surveys being constructed include the SN program for the Vera C. Rubin Observatory\footnote{Formerly known as Large Synoptic Survey Telescope (LSST)} \citep[Rubin/LSST; ][]{2018arXiv180901669T, 2019ApJ...873..111I} and the Nancy Grace Roman Space Telescope\footnote{Formerly known as the Wide-Field InfraRed Telescope (WFIRST)} \citep[Roman/WFIRST; ][]{Rubin2021PASP..133f4001R,2018ApJ...867...23H}.  The Kunlun Dark Universe Survey Telescope \citep[KDUST; ][]{2011PASP..123..725Z, 2014SPIE.9145E..0EZ} and the European Extremely Large Telescope\footnote{https://elt.eso.org/} (ELT) may also have the potential to find and observe the first generation SNe~Ia at the epoch approaching reionization.

The next-generation telescopes like JWST, however, open a new opportunity for cosmological measurements with SNe~Ia. It will be possible to acquire a statistically significant sample out to redshifts up to 6 \citep{Wang2017arXiv171007005W,Wang2019BAAS...51c.399W,Regos2019ApJ...874..158R}. Such data complement those expected from Rubin/LSST and Roman/WFIRST observatories. The more nearby SNe~Ia are from progenitors with more diverse ranges of ages and metallicities; the higher redshift SNe are systematically produced by younger progenitor stars which are also likely of lower metallicities than their lower redshift counterparts. A sample of VHZ SNe~Ia may serve as the cornerstone to build up the framework to quantify the systematic evolution of the physical properties of SNe~Ia.  

The goal of this study is to explore the constraining power of the VHZ SNe~Ia. 
The SNIa data we employ consist of two parts: the comparatively lower-$z$ (CLZ) data from the Pantheon compilation \citep{Scolnic2018ApJ...859..101S} data and a mock VHZ data set based on the capabilities of JWST \citep{Wang2017arXiv171007005W, Regos2019ApJ...874..158R}. 
The cosmological models we employ are the $\Lambda$CDM model, $w$CDM models,  $w_0w_a$CDM and Flat $w_0w_a$CDM models \citep{2001IJMPD..10..213C, 2003PhRvL..90i1301L}. 


The structure of the paper is organized as follows: 
In Section~\ref{sec:model}, we review the basic framework of the cosmological models.
In Section~\ref{sec:SNdata}, we describe the data sample used for constraining cosmological parameters, including the generation of the mock VHZ data sample. The constraints on the various cosmological models are shown in Section~\ref{sec:para}.  A summary of the study is given in Section~\ref{sec:conclusion}.

\section{Cosmological Models} \label{sec:model}
Modern cosmology is built upon the cosmological principle and general relativity \citep[see e.g.][]{Weinberg2008cosmology}. 
In the cosmological models, the luminosity distance $d_L$ can be calculated from
\begin{equation}
    d_L= \frac{c}{H_0}
    {\lim\limits_{{\Omega'_k} \to {\Omega_k}}} 
    \rm{sinh}\left[\sqrt{\Omega'_k}
    \int_1^z\frac{dz'}{E(z')}\right]
\end{equation}

\begin{equation}
    E(z)=\sqrt{\Omega_M(1+z)^3+\Omega_\Lambda(1+z)^{3(1+w)}+\Omega_k(1+z)^2}
\end{equation}
where $H_0$ is the Hubble constant, $w$ describes the equation of state of the dark energy, $\Omega_M$, $\Omega_{\Lambda}$ and $\Omega_{k}$ represent the density parameters of matter, dark energy, and curvature, respectively. 

In the $\Lambda$CDM model, the dark energy has the form of vacuum energy with the equation of state given by $w = -1$. In the $w$CDM model, the dark energy equation of the state of $w$ is also a constant, but it could be different from $-1$. \cite{2001IJMPD..10..213C} and \cite{2003PhRvL..90i1301L} proposed the $w_0w_a$CDM model with $w(z)$ parameterized as
\begin{equation}
    w(z)=w_0+w_a(1-a)
 =w_0+w_a\frac{z}{1+z}
    \label{eq:w0wa}
\end{equation}
When $w_a=0$, the $w_0w_a$CDM model will  recover the $w$CDM model.  When both $w_a=0$ and $w_0=-1$, the $w_0w_a$CDM model will be identical to the $\Lambda$CDM model. The Flat $w_0w_a$CDM model is the $w_0w_a$CDM model that assumes $\Omega_k=0$ i.e. $\Omega_M+\Omega_\Lambda=1$.

\section{Type Ia Supernova Data}\label{sec:SNdata}
  
Instead of working on simulated data achievable with the upcoming telescopes such as Rubin/LSST and Roman/WFIRST observatories, we choose to use the currently available SN data set for the CLZ SNe~Ia. 
The Pantheon compilation is the largest available SNIa data set. It includes 1048 SNe~Ia in a redshift range of $z = 0.010 \sim 2.26$ \citep{Scolnic2018ApJ...859..101S}. The Pantheon compilation consists of data from five different groups: PS1 \citep[Pan-STARRS1,][]{2014ApJ...795...44R}, SDSS \citep[][]{2008AJ....135..338F, 2009ApJS..185...32K}, SNLS \citep[SuperNova Legacy Survey,][]{2011ApJS..192....1C, 2011ApJ...737..102S}, Low-$z$ \cite[CfA,][]{2009ApJ...700.1097H,2009ApJ...700..331H} and \textit{HST} \citep[Union2.1,][]{2012ApJ...746...85S}. The data include the redshifts in the framework of CMB, the distance moduli, and the covariance matrix for the distance moduli\footnote{https://github.com/dscolnic/Pantheon}. For our studies, we will use the redshift binned Pantheon data.
We have tested our results using the binned data and the individual SNIa data and they are consistent in all cases.

\begin{figure}
    \centering
    \includegraphics[scale=0.5]{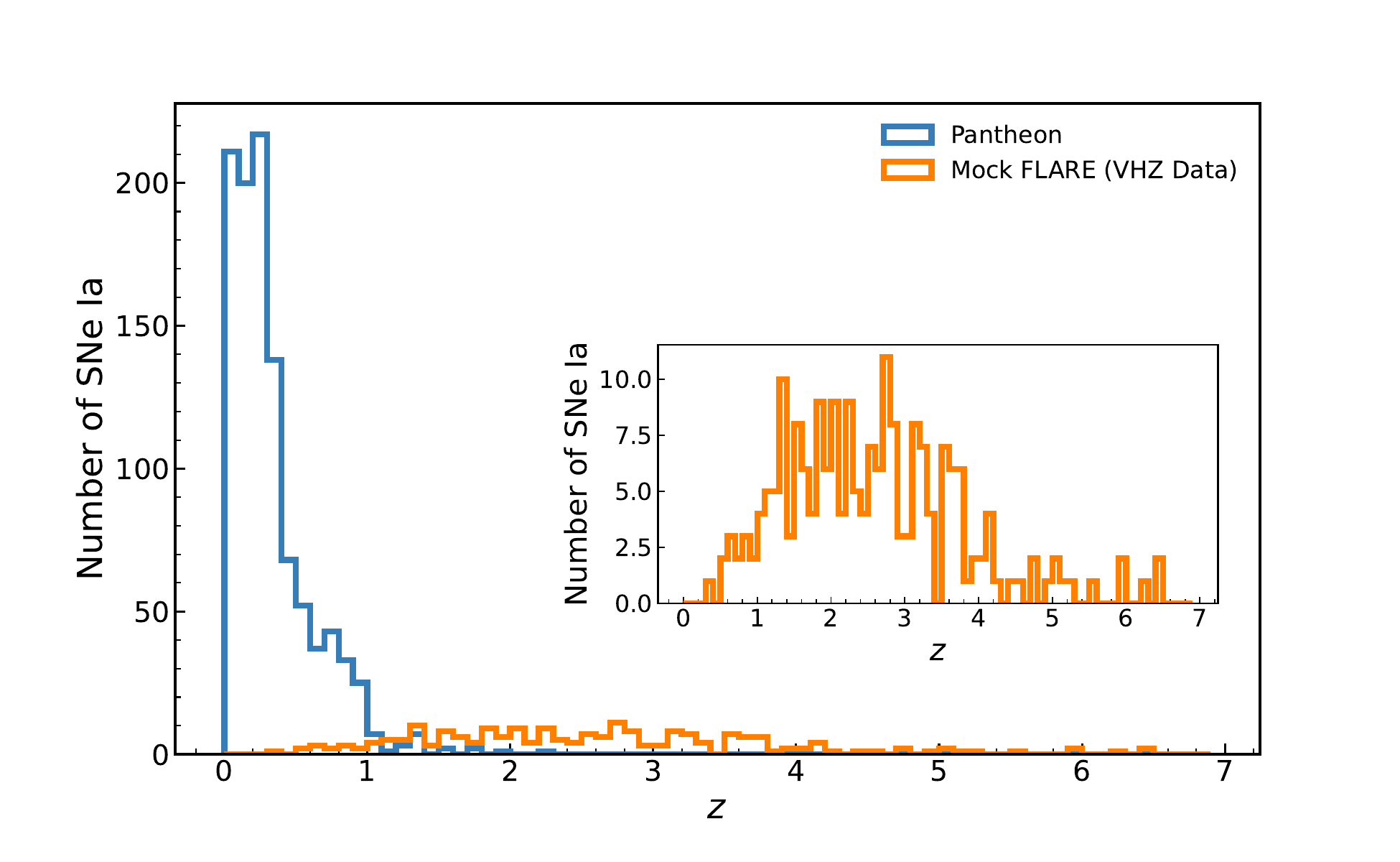}
    \caption{The Redshift distribution of Pantheon SNe~Ia and the mock VHZ Data. 
    }
    \label{fig:SNzNHist}
\end{figure}

\begin{figure}
\centering
\includegraphics[scale=0.5]{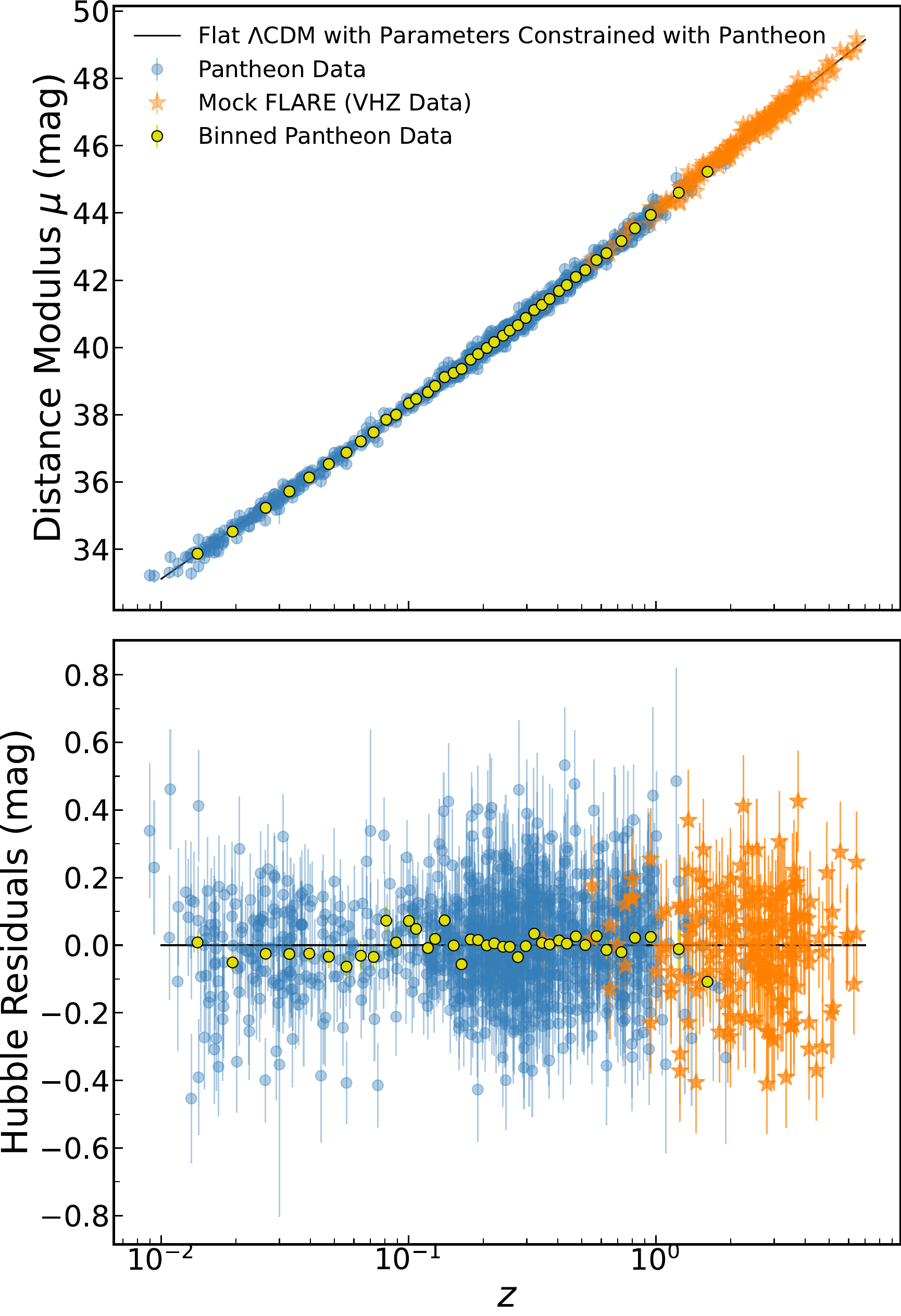}
\caption{Top: Hubble Diagram of the Pantheon unbinned and binned data and the mock VHZ SNe~Ia , and the best fit Flat $\Lambda$CDM model of the Pantheon data.
Bottom: The residuals of the Pantheon and VHZ SNIa data after the subtraction of the best fit Flat $\Lambda$CDM cosmology.}
\label{fig:PanJWSTHD}
\end{figure}

The redshifts of the VHZ SNe~Ia are derived following the FLARE project \citep{Wang2017arXiv171007005W, Regos2019ApJ...874..158R} where the rates of SNIa are based on an extrapolation of the local SNIa rates out to z $\sim$ 6 based on the star formation rates at higher redshifts as recently derived in \cite{ChenXingzhuoMuse2021arXiv210106242C}. Note that the rates given in \cite{ChenXingzhuoMuse2021arXiv210106242C} are in the observer time frame. The survey assumes an area of 0.05 square degrees with a time span of 6 years.  In practice, the VHZ SN survey can be divided into two identical stages, each of 3 years duration with a cadence of 91 days to fully match the footprints allowed by the JWST NIRCAM observations at each observing visit. The proposed survey may employ four broadband JWST NIRCAM filters (F150W, F200W, F322W2, and F444W) with exposure times that can reach 10 $\sigma$ limiting magnitudes of 27.7, 27.7, 28.1, and 27.2 mag in these filters, respectively. The regions in the JWST Continuous Viewing Zone (CVZ) are chosen for the potential survey field, as suggested by the PEARLS team for their time domain survey for the JWST \citep{JansenTDF2018PASP..130l4001J}. The multicolor light curves can ensure the robust classification of the SNe~Ia. A small fraction of the SNe~Ia will be observed spectroscopically with the JWST NIRSpec for studies of the spectral properties of these SNe. In most cases, only single-epoch spectroscopic data will be attempted. Recent studies based on well-observed nearby SNe~Ia show that a single epoch spectrum of an SN~Ia around optical maximum can allow for accurate reconstruction of the spectral sequence of the SN from two weeks before to one month after the optical maximum \citep{HuLSTM2022ApJ...930...70H}. A single epoch spectrum can be sufficient in fully defining the observational properties of an SN~Ia if the VHZ SNe are members of the sub-Types observed already by the nearby SN surveys. Multiple epoch spectra may also be requested to test the hypothesis that the nearby SNe contains events that are similar to the VHZ SNe; this offers a crucial examination of the redshift evolution of the intrinsic properties of SNe~Ia out to the first generation SNe~Ia of the Universe.

We simulated the occurrence of the SNe~Ia in such a survey strategy using the Monte Carlo method and found that the total number of SNe~Ia expected in a 6-year survey is of the order of 200.  We will base our study in this paper on a  particular realization of the simulation, which yields a total of 205 SNe~Ia.

The mock VHZ SNIa sample is constructed based on a Flat $\Lambda$CDM fit to the Pantheon data to ensure consistency between the local and distant samples. We assume that 75\% of all the detectable SNe~Ia will lead to reliable measurements of luminosity distances to the host galaxies of the SNe.
The luminosity distances of the mock SNe are calculated using the mock SN redshifts 
and the Flat $\Lambda$CDM model with $H_0 = 71.66\ \rm km \cdot s^{-1}\cdot Mpc^{-1}$ and $\Omega_M = 0.30$ as derived from a Flat $\Lambda$CDM fit to the Pantheon data. 
A Gaussian error of 0.15 mag, which represents typical distance measurement precision achievable by SNe~Ia \citep[e.g.][]{1993ApJ...413L.105P}, is added to the distance moduli of the SNe~Ia to account for the statistical uncertainties of distance determinations. This prescription of the error is an oversimplification, and realistic distance errors depend on the details of the observational setup. Studies of SNe~Ia as distance indicators have demonstrated that much smaller Hubble residuals of 0.07 mag can be achieved in some statistical methods \citep{WangCMAGIC2003ApJ...590..944W, WangStrovink2006,He2018ApJ...857..110H,TwinsEmbeddingI2021ApJ...912...70B, TwinsEmbeddingII2021ApJ...912...71B}. In particular, the Twins Embedding method, which matches spectral features of SNe~Ia, is extremely promising in significantly reducing both the statistical errors and any potential systematic redshift evolutions of the SNIa luminosity distance measurements \citep[][]{NearbySupernovaFactory:2021afn, Boone_2021}. For the covariance matrix, we assume that all the off-diagonal elements involving mock SNe~Ia are given by $\hat{C}=0.002^2$ with a random $+$ or $-$ sign assigned to each upper right diagonal component and mirrored to the lower diagonal to keep the covariance matrix symmetric. The value $0.002^2$ is equal to the median value of the off-diagonal components of the covariance matrix of the Pantheon compilation. The distance moduli of the mock SNIa sample are generated under the assumption of this amount of covariant errors using the covariance matrix of the entire data set for consistency. We do not assume any redshift dependence of the covariance matrix in this study. 

A histogram of the redshifts of this data set is displayed in Figure~\ref{fig:SNzNHist}. The Hubble diagram 
and residuals of the combined Pantheon and mock VHZ SNIa data are shown in Figure~\ref{fig:PanJWSTHD}. 

\section{Cosmological Parameters and Systematic Effects}\label{sec:para}

The ultimate goal of model selection and parameter estimation in cosmology is to find the best single model and the values of the model parameters to describe the expansion of the universe.
Here we will focus on the difference between the fits with and without the mock VHZ SNIa data. 

The most significant impact on cosmological parameter estimation by the addition of the VHZ SNIa data set is that it enables systematic redshift evolution of the SNIa luminosity distances to be explicitly studied throughout the entire stellar evolution history of the universe. This makes it possible to disentangle complex theoretically expected evolutionary effects related to the age and metallicity of the progenitor systems from the effects due to the underlying cosmology. The error arising from extinction corrections due to the systematic evolution of the interstellar dust properties for host galaxies at different redshifts is a well-known issue and remains difficult to properly quantify \citep[see e.g.][]{WangStrovink2006}. The physical origins of the environmental effect such as the host galaxy mass step
dependence \citep{MassStepSullivan2010MNRAS.406..782S,MassKelly2010, UddinMass2017ApJ...848...56U,MassStep2020ApJ...901..143U, MassStep2021arXiv210506236J} are poorly understood. The mass step dependence itself is expected to be redshift dependent since the mean mass of SNIa producing galaxies is redshift dependent. As shown by \cite{Rigault2020A&A...644A.176R}, approximately 70\% of the variance from the stellar mass step is due to an underlying dependence on environment-based progenitor age, and the local specific star formation rate within a projected distance of 1 kpc is a better indicator than the mass step. Furthermore, \cite{UddinSpec2017ApJ...850..135U} also found that the average spectral properties of the SNIa hosts are different for SNe~Ia with different light curve decline rates past optical maximum. \cite{WangHost} found that the intrinsic magnitude dispersion of SNe~Ia after light curve shape and color corrections show systematic dependence on the galaxy-centric distances of the SNe~Ia and the effect born out by recent data \citep{MassStep2020ApJ...901..143U}. From a theoretical point of view, it is also likely that both the single degenerate (SD) and double degenerate (DD) models may be at work in the production of SNe~Ia. In addition to the popular SD scenario, recent studies seem to indicate that most of the observed SNe~Ia can be accounted for by a delay time distribution (DTD) that can be expected from the DD channel \citep{CastrilloMuse10.1093/mnras/staa3876,ChenXingzhuoMuse2021arXiv210106242C,WisemanRates2021MNRAS.506.3330W}. \cite{WisemanRates2021MNRAS.506.3330W} found that there is a strong correlation between the DTD slope and the SN light curve decline rate. The SNIa rates were also found to be different for SNe~Ia with different color parameters as measured from their light curves. Considering that the light curve decline rate and the color parameter are the two most critical parameters in SNIa distance standardization, it is thus critically important to explore mechanisms that can quantitatively probe the systematic redshift dependence of standardized SNIa luminosity distances. 

Along with the underlying cosmology, physical processes such as the magnification/de-magnification of the SN magnitudes by large-scale structures \citep{WambsganssLensing1997,Jonsson:LensingSNLS, WangYun1999ApJ...525..651W, SakakibaraLensing10.1093/mnras/stz1117} may also introduce magnitude evolution that cannot be accounted for by SN standardization methods. The magnification probability is a function of redshift. At increasing redshift the most likely magnification factor shifts to lower values and the dispersion of the probability density distribution of magnification increases \citep{WambsganssLensing1997,WangYun1999ApJ...525..651W}. At $z \sim 2$ and $6$, the peak of the probability of the lensing magnification shifts to $\sim 0.95$, and the dispersion of the magnification around the peak doubles from $0.05$ mag at $z\sim 2$ to $0.10$ mag at $z\sim 5$ \citep{WangYun1999ApJ...525..651W}. An increasingly larger number of highly magnified SNe~Ia will be picked up by magnitude-limited surveys than their de-magnified counterparts, leading to a systematic bias that needs to be considered carefully, especially when the VHZ SNe~Ia are employed for cosmological constraints. However, it should also be noted that the gravitational lensing magnification itself is strongly dependent on cosmological parameters. A positive detection of the lensing effect from the VHZ SNe~Ia may lead to independent constraints on cosmological models. 

The best use of the VHZ SNIa will certainly include comparisons of the light curves and spectra with their local 
counterparts as in \cite{TwinsEmbeddingI2021ApJ...912...70B, TwinsEmbeddingII2021ApJ...912...71B}. Data-driven analyses of SNIa observations and theoretical models \citep{ChenXingzhuoMuse2021arXiv210106242C} may also enable quantitative derivation of physical parameters that can be used to reduce systematic errors. However, it is instructive to understand what statistical capabilities the VHZ SNe~Ia may empower based on the simplest statistical approach.
A number of systematic effects will be investigated in this study with the mock VHZ SNIa data. First, to account for a possible systematic luminosity difference between the VHZ SNe~Ia and CLZ SNe~Ia, we introduce a magnitude bias between the VHZ SNe~Ia and the CLZ SNe~Ia. Mathematically, a bias level $\hat{\Gamma}_0$ is added to the luminosity distances of the VHZ SNe~Ia, and $\hat{\Gamma}_0$ is taken as a free nuisance parameter in the model fits to constrain cosmological models. 
Secondly, for the intrinsic evolution of the standardized SNIa luminosity distance with redshift, we will assume a linear redshift dependence of magnitude evolution, that is, $\delta\mu(z)  = \hat{\Gamma}_1 z$ and a logarithmic evolution given by $\delta\mu(z) = \hat{k}\ln(1+z)$ \citep{2000ApJ...530..593D}. The parameters $\hat{\Gamma}_1$ and $\hat{k}$ will be taken as free parameters to be constrained by the SNIa data.  In principle, the maximum amount of magnitude evolution can be constrained with the residuals of the Hubble diagrams constructed using the CLZ SNe~Ia. We have learned from observations of local SNe~Ia that the intrinsic dispersion of the SNe~Ia after light curve and color corrections is around 0.15 magnitudes \citep{Riess1999AJ....117..707R,Perlmutter1999ApJ...517..565P,Scolnic2018ApJ...859..101S}, which can be used as a prior on the systematic evolution.
The SNe~Ia that are analyzed for cosmological studies represent only the normally behaving subset of the entire SNIa sample; some nearby SNe~Ia do show  Hubble residuals larger than 0.15 mag. There is no guarantee that peculiar subgroups will not grow in importance at increasingly higher redshifts, especially for future Rubin/LSST-related SN surveys relying heavily on photometric SN classifications. We do not know whether the SNe~Ia with large Hubble residuals will be more frequently encountered at VHZ. We thus will set no prior on the coefficients $\hat{\Gamma}_1$ and $\hat{k}$ in this study.

\begin{figure}
\includegraphics[width=0.4\textwidth]{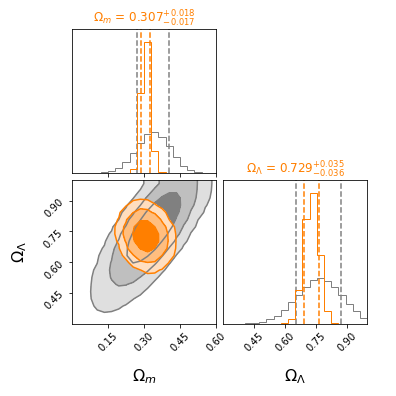}
\includegraphics[width=0.6\textwidth]{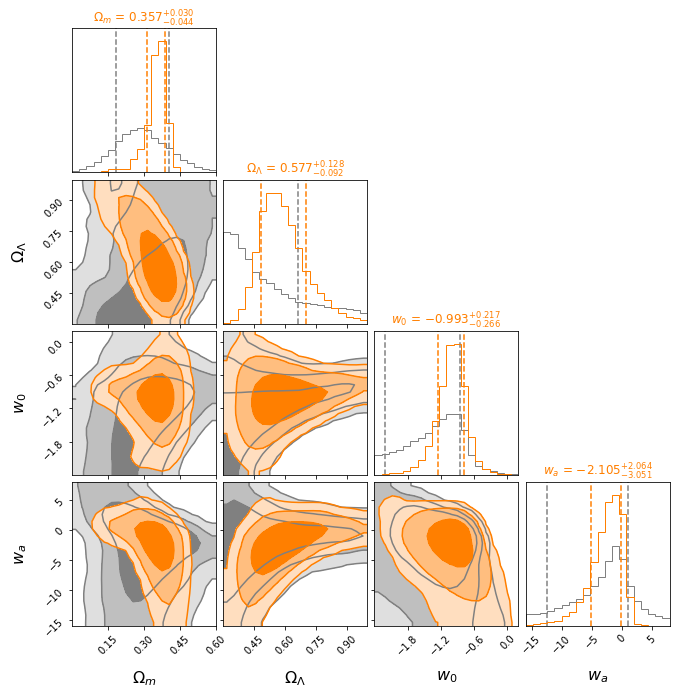}
\includegraphics[width=0.5\textwidth]{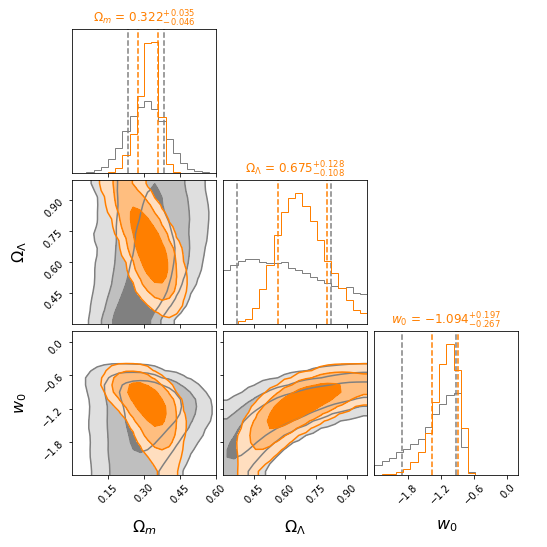}
\includegraphics[width=0.5\textwidth]{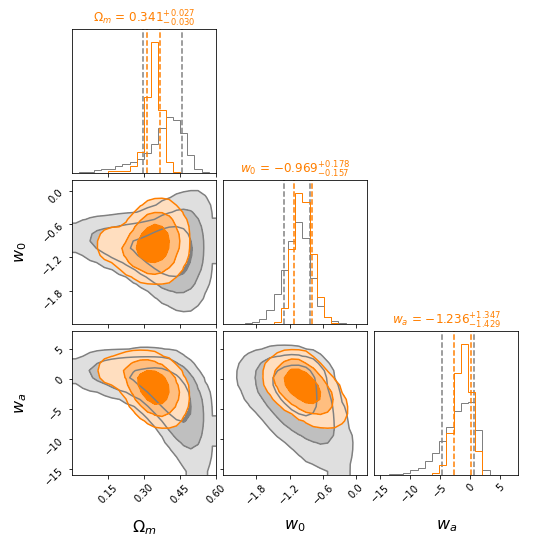}
\caption{Clockwise from upper left, the contours showing cosmological constraints with (orange) and without (gray) the VHZ SNIa data, for the $\Lambda$CDM, $w_0w_a$CDM, and Flat $w_0w_a$CDM, $w$CDM. The titles in each panel show the central values of the cosmological parameters and their 1$\sigma$ uncertainties for the joint Pantheon and VHZ SNIa data set. The contour levels are 1, 2, and 3-$\sigma$ with decreasing color weight.}
\label{fig:NoEvol}
\end{figure}

To derive the optimal constraints on the cosmological parameters, we adopt the Markov Chain Monte Carlo (MCMC) code emcee \citep{emcee2013PASP..125..306F} for model parameter calculations.
The expression of $\chi^2$ with statistical and systematic errors to be minimized is thus set as
\begin{equation}\label{eq:chi2}
\chi^2=\sum\limits_{i=0}^{N-1}\sum\limits_{j=0}^{N-1}  \Delta \mu({z_i}) 
 C_{ij}^{-1}\Delta \mu({z_j}) 
\end{equation}
where $\Delta\mu\ =\ \mu_{\rm{data}}(z)-\mu_{\rm{model}}(z; \vec{P})-\hat{\Gamma}_0 B(z)-\delta\mu(z)$, with $B(z)$ being a function that is equal to 1 for the VHZ SNe~Ia but 0 otherwise, $\mu_{\rm{data}}(z)$ the distance modulus of the SNIa data, $\mu_{\rm{model}}(z; \vec{P})$ the corresponding distance modulus for a particular cosmological model with parameter set ${\vec{P}}$, and $\delta\mu(z)$ is the magnitude evolution defined above. 

With the above equation, two more parameters are introduced in fitting the SNIa data, these are $\hat{\Gamma}_0$ and $\hat{\Gamma}_1$, or $\hat{\Gamma}_0$ and $\hat{k}$. Note that $\hat{\Gamma}_0$ only accounts for the systematic magnitude offset between the VHZ SNe~Ia and the local SNe~Ia, and is not needed if the distance scales of the two sets of data are calibrated precisely with no systematic differences. Our aim is to explore how the VHZ SNe~Ia may set constraints on the function $\delta\mu(z)$. 
For the different cosmological models we have $\vec{P} = (\hat{H}_0, \Omega_M, \Omega_\Lambda)$, $\vec{P} = (\hat{H}_0, \Omega_M, \Omega_\Lambda, w_0)$, $\vec{P} = (\hat{H}_0, \Omega_M, \Omega_\Lambda, w_0, w_a)$, and $\vec{P} = (\hat{H}_0, \Omega_M, w_0, w_a)$ for the $\Lambda$CDM, $w$CDM, $w_0w_a$CDM, and Flat $w_0w_a$CDM model, respectively. Note that $\hat{H}_0$ is not really an estimate of the Hubble constant, but a nuisance parameter that is related to the absolute magnitude of the SNe~Ia after standardization.  

\subsection{Models with No Systematic Redshift Evolution of the Intrinsic Properties of SNe~Ia}\label{subsection:NoEvol}

\begin{table}[]
    \centering
    \caption{Constraints on Cosmological Parameters of  $\Lambda$CDM}
    \begin{tabular}{l|cccccc}
    \hline
    \hline 
    Dataset   & $\Omega_m$ & $\Omega_\Lambda$ & $\hat{\Gamma}_0$ & $\hat{\Gamma}_1$ & $\hat{k}$\\
\hline
Pan\tablenotemark{a}   & $0.338_{-0.068}^{+0.068}$ & $0.766_{-0.111}^{+0.105}$  & & \\
Pan+P\tablenotemark{a}   & $0.272_{-0.003}^{+0.003}$ & $0.753_{-0.006}^{+0.006}$  & & \\
PanVHZ\tablenotemark{a}  & $0.307_{-0.017}^{+0.018}$& $0.729_{-0.036}^{+0.035}$&
$0.040_{-0.028}^{+0.028}$ &  & \\
PanVHZ+P\tablenotemark{a}  & $0.273_{-0.003}^{+0.003}$ & $0.755_{-0.006}^{+0.006}$  &$-0.013_{-0.012}^{+0.012}$ & & \\
\hline
Pan\tablenotemark{b} &  $0.559_{-0.284}^{+0.278}$& $0.765_{-0.125}^{+0.117}$&&
$0.185_{-0.242}^{+0.170}$ &  \\
Pan+P\tablenotemark{b} &  $0.275_{-0.004}^{+0.004}$ & $0.745_{-0.008}^{+0.008}$  & &$-0.042_{-0.024}^{+0.024}$ &  \\
PanVHZ\tablenotemark{b}& $0.334_{-0.063}^{+0.065}$ & $0.747_{-0.051}^{+0.051}$  & $0.046_{-0.031}^{+0.032}$ & $0.017_{-0.039}^{+0.035}$ & \\
PanVHZ+P\tablenotemark{b} & $0.273_{-0.003}^{+0.003}$ & $0.750_{-0.006}^{+0.006}$ & $0.019_{-0.023}^{+0.022}$ & $-0.014_{-0.009}^{+0.008}$ & \\
\hline
Pan\tablenotemark{c}&  $0.676_{-0.291}^{+0.228}$ & $0.586_{-0.190}^{+0.188}$ & & &
$0.456_{-0.359}^{+0.206}$ \\
Pan+P\tablenotemark{c}&  $0.276_{-0.004}^{+0.004}$ & $0.744_{-0.008}^{+0.008}$  &   & &
$-0.055_{-0.033}^{+0.033}$ \\
PanVHZ\tablenotemark{c} & $0.490_{-0.166}^{+0.232}$ & $0.704_{-0.043}^{+0.045}$  & $0.056_{-0.030}^{+0.031}$  & & $0.226_{-0.201}^{+0.213}$ \\
PanVHZ+P\tablenotemark{c} & $0.275_{-0.003}^{+0.003}$ & $0.746_{-0.008}^{+0.007}$  & $0.033_{-0.027}^{+0.027}$  & & $-0.044_{-0.024}^{+0.023}$ \\
    
\hline
\end{tabular}
   \begin{itemize}
    \item[] \tablenotetext{}{
      Notes to Table entries: Pan stands for Pantheon only, Pan+P for Pantheon data with Planck prior, PanVHZ for Pantheon data and the VHZ SNe~Ia, and PanVHZ+P for Pantheon and VHZ SNe~Ia with Planck prior. }
     \item[] \tablenotetext{a}{No systematic evolution.}
    \item[] \tablenotetext{b}{Linear systematic evolution proportional to $\hat{\Gamma}_1$z.}
    \item[] \tablenotetext{c}{Logarithmic systematic evolution proportional to $\hat{k}\ln(1+z)$.}
    \end{itemize}
   \label{tab:paramsLCDM}
\end{table}

We consider first the cosmological fits with no systematic evolution due to the unknown intrinsic evolution of the SNe~Ia across the cosmic time from $z\sim 0$ to $6$. The systematic errors of the Pantheon binned data and their correlations with the VHZ SNe~Ia are included in these fits as described in Section~\ref{sec:SNdata}. This represents an optimistic case in which the VHZ SNe~Ia may improve the constraints on cosmological parameters. 

As shown in Figure~\ref{fig:NoEvol}, the addition of the VHZ SNe~Ia leads to a significant improvement in the confidence contours to the value of $\Omega_M$ for all the cosmological models. Figure~\ref{fig:NoEvol} shows that the 68\%, 95\%, and 99.7\% confidence levels of $\Omega_M$ and $\Omega_\Lambda$. These two parameters are strongly correlated for the data set with redshifts from 0 to about 2, such as the Pantheon data set. It is seen that the constraints on the parameters related to the dark energy also show significant improvements when $\Omega_M$ is better constrained by the VHZ SNe. 

The parameters involved in the $\Lambda$CDM and Flat $w_0w_a$CDM are shown in Tables~\ref{tab:paramsLCDM} and \ref{tab:paramsFlatw0waCDM} for the $\Lambda$CDM and Flat $w_0w_a$CDM, respectively, constrained by the Pantheon only data (data set Pan), Pantheon only and Planck prior combined (data set Pan+P), Pantheon and VHZ SNe combined (data set PanVHZ), and Pantheon, VHZ SNe and Planck data combined (data set PanVHZ+P), for cases with no systematic evolution, a linear redshift dependent systematic evolution, and a logarithmic systematic evolution. 
As shown in the first and third rows of Table~\ref{tab:paramsLCDM}, for the $\Lambda$CDM model, it is remarkable that the introduction of the VHZ SNe Ia results in a reduction of the sizes of the 1$\sigma$ errors of $\Omega_m$ and $\Omega_\Lambda$ to $\pm0.018$ and $\pm0.036$ from $\pm0.068$ and $\pm0.111$, respectively. For $\Omega_M$ and $\Omega_\Lambda$ in the $\Lambda$CDM
model, 
this improvement is more than a factor of 9 in terms of the area of the error contours. The introduction of the VHZ SNe~Ia changes the direction of the orientation of the confidence contours for the $\Lambda$CDM model, making it nearly vertical to the $\Omega_M$ axis (Figure~\ref{fig:NoEvol}, upper left). 
Similar dramatic improvements are also seen for the $w$CDM, $w_0w_a$CDM, and Flat $w_0w_a$CDM models. For example, as shown in Figure~\ref{fig:NoEvol} (lower right) for the Flat $w_0w_a$CDM, the VHZ SNe Ia when combined with existing Pantheon data, will allow $w_0$ and $w_a$ to be determined to $w_0 = -0.969^{+0.178}_{-0.157}$ and $w_a = -1.236^{+1.347}_{-1.429}$. 
The errors are smaller than any of the Stage III and the pessimistic Stage IV supernova experiments shown in the Dark Energy Task Force (DETF) report \citep{DETF2006astro.ph..9591A}.  

\begin{table}[]
\caption{Constraints on Cosmological Parameters of  Flat $w_0w_a$CDM}
\begin{tabular}{l|ccccccc}
\hline
\hline 
Dataset  & $\Omega_m$ & $w_0$ &$w_a$ & $\hat{\Gamma}_0$ & $\hat{\Gamma}_1$ & $\hat{k}$\\
\hline
Pan\tablenotemark{a}   & $0.397_{-0.100}^{+0.060}$ & $-1.052_{-0.250}^{+0.217}$  & $-1.640_{-3.171}^{+2.219}$ &  \\
Pan+P\tablenotemark{a}   & $0.280_{-0.004}^{+0.005}$ & $-0.847_{-0.088}^{+0.091}$ &
$-0.900_{-0.348}^{+0.322}$ & \\
PanVHZ\tablenotemark{a}  & $0.341_{-0.030}^{+0.027}$ & $-0.969_{-0.157}^{+0.178}$ &$-1.236_{-1.429}^{+1.347}$ & $0.040_{-0.033}^{+0.033}$ & \\
PanVHZ+P\tablenotemark{a}  & $0.279_{-0.005}^{+0.005}$ & $-0.887_{-0.088}^{+0.088}$&
$-0.776_{-0.336}^{+0.323}$&
$-0.038_{-0.011}^{+0.011}$&
\\
\hline
Pan\tablenotemark{b} &  
$0.666_{-0.163}^{+0.105}$ & $-1.324_{-0.718}^{+0.570}$  & $4.087_{-12.382}^{+4.504}$ & & $0.401_{-0.227}^{+0.341}$& \\
Pan+P\tablenotemark{b} &  
$0.279_{-0.005}^{+0.005}$ & $-1.010_{-0.145}^{+0.147}$  & $-0.282_{-0.548}^{+0.509}$ & & $-0.055_{-0.040}^{+0.039}$& \\
PanVHZ\tablenotemark{b}& $0.378_{-0.064}^{+0.053}$ & $-0.965_{-0.192}^{+0.200}$  & $-1.881_{-2.189}^{+1.803}$ &  $0.047_{-0.034}^{+0.034}$ &  $0.014_{-0.019}^{+0.016}$\\
PanVHZ+P\tablenotemark{b} & $0.279_{-0.004}^{+0.004}$ & $-0.910_{-0.087}^{+0.088}$  & $-0.664_{-0.330}^{+0.315}$ & $0.008_{-0.023}^{+0.023}$ & $-0.019_{-0.008}^{+0.008}$\\
\hline
Pan\tablenotemark{c}&  $0.799_{-0.158}^{+0.097}$ & $-1.328_{-0.985}^{+0.923}$  & $-0.009_{-17.052}^{+6.222}$   & & & $0.693_{-0.220}^{+0.252}$ \\
Pan+P\tablenotemark{c}&  
$0.279_{-0.005}^{+0.005}$ & $-1.039_{-0.169}^{+0.165}$  & $-0.167_{-0.622}^{+0.585}$   & & & $-0.082_{-0.063}^{+0.060}$ \\
PanVHZ\tablenotemark{c} & $0.503_{-0.136}^{+0.302}$ & $-0.838_{-0.308}^{+0.431}$  & $0.296_{-4.296}^{+3.293}$  &$0.042_{-0.035}^{+0.034}$ & & $0.191_{-0.144}^{+0.785}$ \\
PanVHZ+P\tablenotemark{c}&  $0.279_{-0.004}^{+0.005}$ & 
$-0.989_{-0.093}^{+0.098}$  & $-0.370_{-0.361}^{+0.334}$  & $0.027_{-0.028}^{+0.028}$ & & $-0.057_{-0.023}^{+0.023}$ \\
\hline
\end{tabular}
\begin{itemize}
\small{
\item[] \tablenotetext{}{Notes to Table entries: Pan stands for Pantheon only, Pan+P for Pantheon data with Planck prior, PanVHZ for Pantheon data and the VHZ SNe~Ia, and PanVHZ+P for Pantheon and VHZ SNe~Ia with Planck prior. }
\item[] \tablenotetext{a}{No systematic Evolution.}
\item[] \tablenotetext{b}{Linear systematic Evolution proportional to $\hat{\Gamma}_1$z.}
\item[] \tablenotetext{c}{Logarithmic systematic Evolution proportional to $\hat{k}\ln(1+z)$.}}
\end{itemize}
\label{tab:paramsFlatw0waCDM}
\end{table}

\begin{figure}
\includegraphics[width=0.5\textwidth]{LambdaCDMnoevolbiasnoCMBtimes1binnedOmOde.png}
\includegraphics[width=0.5\textwidth]{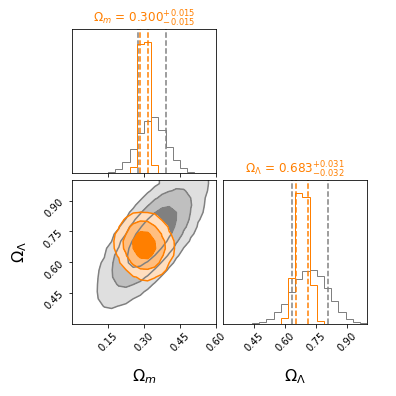}
\includegraphics[width=0.5\textwidth]{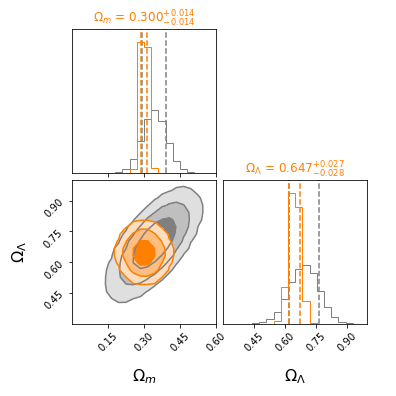}
\includegraphics[width=0.5\textwidth]{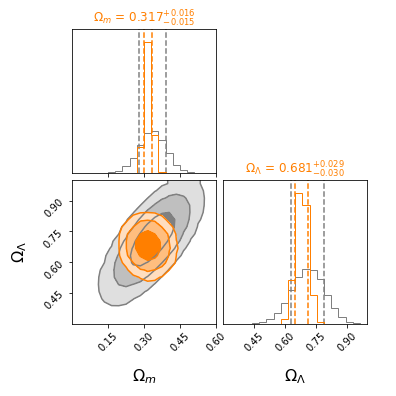}
\caption{From upper-left clockwise, the panels show the constraints on $\Omega_m$ and $\Omega_\Lambda$ of the $\Lambda$CDM model with the statistical errors of the Pantheon binned data reduced by a factor of 1, $\sqrt{8}$, 4, and 8. The gray contours show the confidence levels constrained by the Pantheon binned data only, and the orange contours show those with the inclusion of the VHZ SNe Ia. The gray and orange vertical dashed lines show the 1$\sigma$ confidence levels constrained by the Pantheon binned data only and those of the combined data, respectively. The numbers of the image titles show the 1$\sigma$ confidence of the combined data. The contour levels are 1, 2, and 3-$\sigma$ with decreasing color weight.}
\label{fig:NoEvolLCDM}
\end{figure}

\begin{figure}
\includegraphics[width=0.5\textwidth]{Flatw0waCDMnoevolbiasnoCMBtimes1binnedOmw0wa.png}
\includegraphics[width=0.5\textwidth]{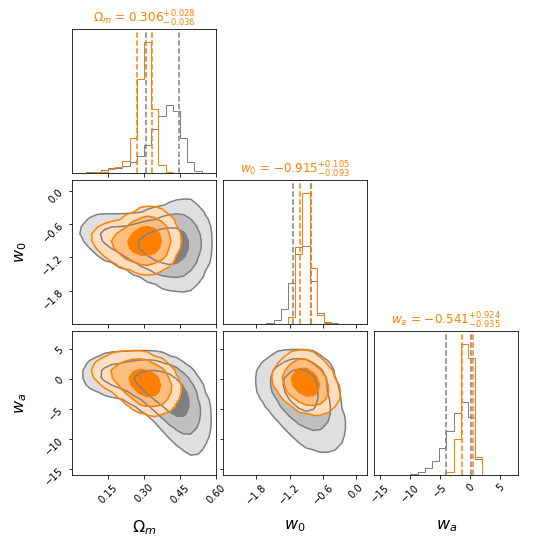}
\includegraphics[width=0.5\textwidth]{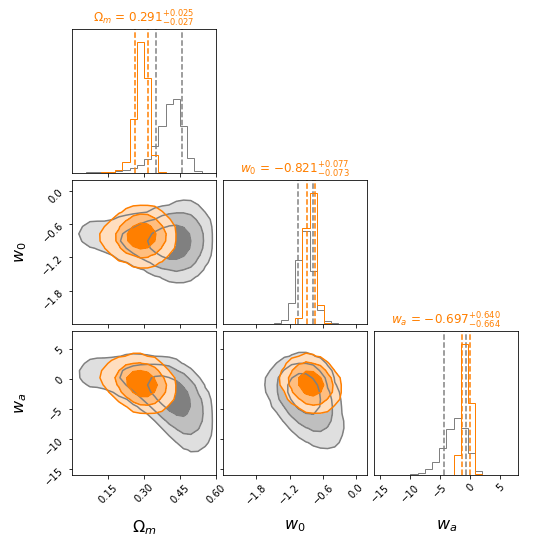}
\includegraphics[width=0.5\textwidth]{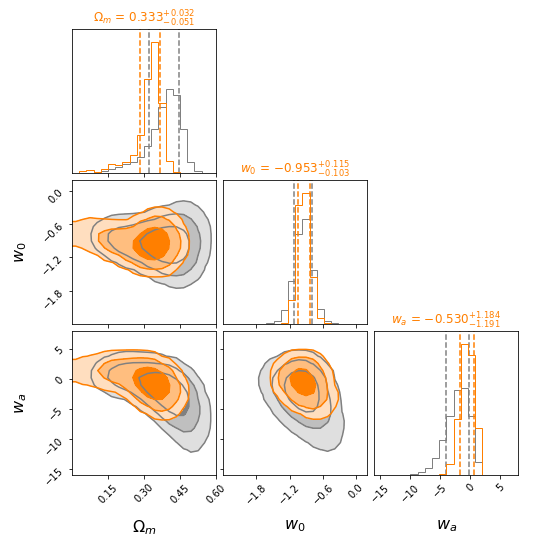}
\caption{From upper-left clockwise, the panels show the constraints on $\Omega_m$, $w_0$ and $w_a$ of the Flat $w_0w_a$CDM model with the statistical errors of the Pantheon binned data reduced by a factor of 1, $\sqrt{8}$, 4, and 8. The gray contours show the confidence levels constrained by the Pantheon binned data only; the orange contours show those with the inclusion of the VHZ SNe Ia. The gray and orange vertical dashed lines show the 1$\sigma$ confidence levels of the Pantheon binned data only and those of the combined data, respectively. The numbers of the image titles show the 1$\sigma$ confidence of the combined data. The contour levels are 1, 2, and 3-$\sigma$ with decreasing color weight.}
\label{fig:NoEvolw0waCDM}
\end{figure}

\begin{figure}
\includegraphics[width=0.4\textwidth]{LambdaCDMnoevolbiasnoCMBtimes8binnedOmOde.png}
\includegraphics[width=0.6\textwidth]{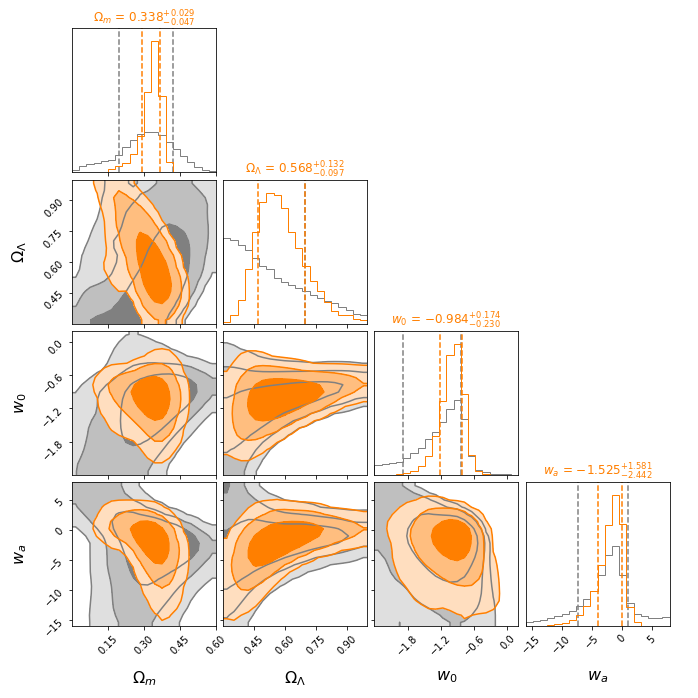}
\includegraphics[width=0.5\textwidth]{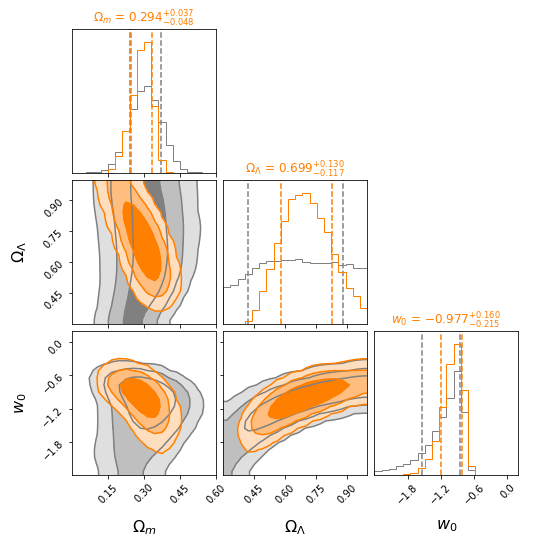}
\includegraphics[width=0.5\textwidth]{Flatw0waCDMnoevolbiasnoCMBtimes8binnedOmw0wa.png}
\caption{Same as Figure~\ref{fig:NoEvol}, but the statistical errors of the lower redshift SN sample are reduced artificially by a factor of $\sqrt{8}$. }
\label{fig:NoEvol8}
\end{figure}

\begin{figure}
\includegraphics[width=0.4\textwidth]{LambdaCDMnoevolbiasnoCMBtimes16binnedOmOde.png}
\includegraphics[width=0.6\textwidth]{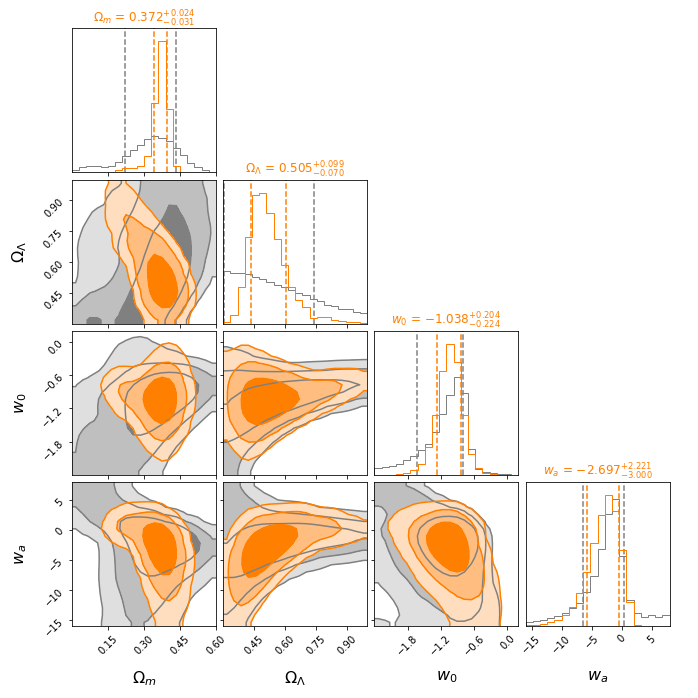}
\includegraphics[width=0.5\textwidth]{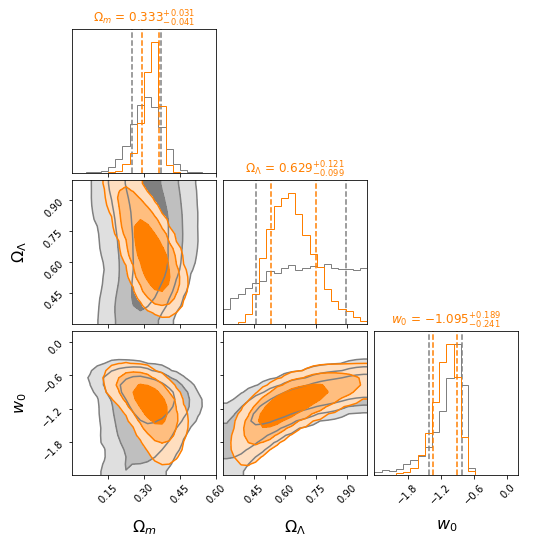}
\includegraphics[width=0.5\textwidth]{Flatw0waCDMnoevolbiasnoCMBtimes16binnedOmw0wa.png}
\caption{Same as Figure~\ref{fig:NoEvol} and Figure~\ref{fig:NoEvol8}, but the statistical errors of the lower redshift SN sample are decreased artificially by a factor of 4. 
\label{fig:NoEvol16}}
\end{figure}

Future supernova surveys with the Rubin/LSST and the Roman/WFIRST observatories may provide orders of magnitude more well observed SNe~Ia. 
Thus it is instructive to see how a decrease of the statistical errors of the Pantheon binned data may improve the results of the cosmological fits. From Figure~\ref{fig:NoEvol} we found that the VHZ SNeIa can add new power to cosmological parameter measurements. A related question is whether a reduction of the statistical errors of the existing binned Pantheon data can further tighten the error contours. Such a data set can be obtained by the continuation of similar observing programs using existing facilities. To do this, the statistical errors of the Pantheon data are reduced by an arbitrary factor $\sqrt{N}$, which is equivalent to increasing the Pantheon data sample by a factor of $N$. The comparisons are shown in Figure~\ref{fig:NoEvolLCDM} for the $\Lambda$CDM model and in Figure~\ref{fig:NoEvolw0waCDM} for the Flat $w_0w_a$CDM model. For both cosmological models, the constraints from the Pantheon data alone show only moderate improvements with the reduction of the statistical noise. The same phenomenon can also be seen for the $w$CDM and the $w_0w_a$CDM model at Figure~\ref{fig:NoEvol8} and Figure~\ref{fig:NoEvol16}. This suggests that the confidence levels have reached their systematic error limits. The inclusion of the VHZ SNe~Ia changes this significantly. Improvements are observed for the $\Lambda$CDM and the Flat $w_0w_a$CDM out to $N = 64$. For the Flat $w_0w_a$CDM,  the constraints on $w_a$ drop below $\pm0.664$.  . This represents the most optimistic observing strategy leading to what current data can achieve when combined with the VHZ SNe~Ia. Further improvement will rely on programs with even more stringent control of systematic errors.

\begin{figure}
\includegraphics[width=0.5\textwidth]{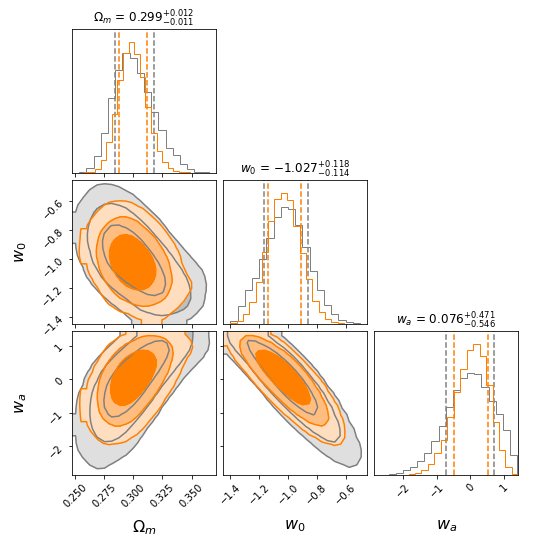}
\includegraphics[width=0.5\textwidth]{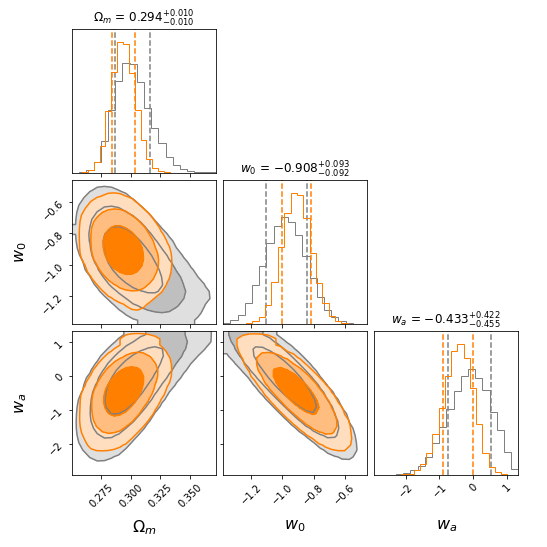}
\includegraphics[width=0.5\textwidth]{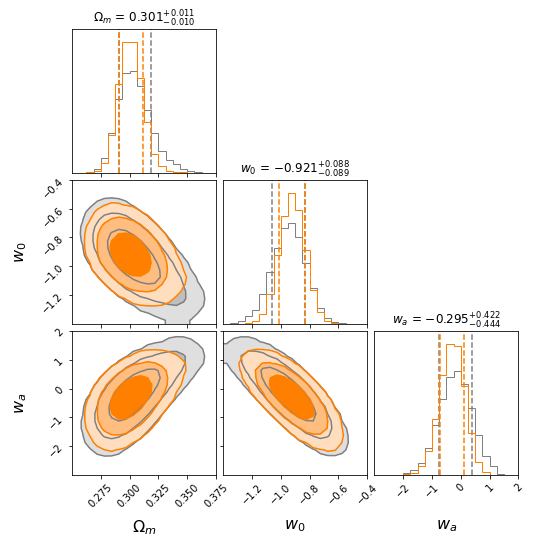}
\includegraphics[width=0.5\textwidth]{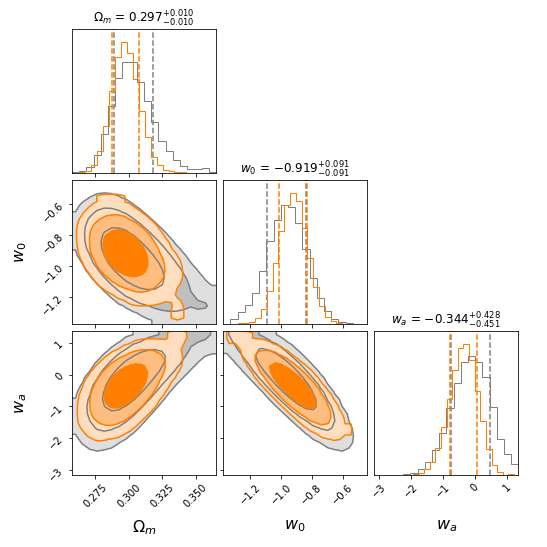}
\caption{From upper-left clockwise, the panels show the constraints on $\Omega_m$, $w_0$, and $w_a$ of the Flat $w_0w_a$CDM model with the Planck 2018 measurements and the statistical errors of the Pantheon binned data reduced by a factor of 1, 2, $\sqrt{8}$ and 8. The gray contours show the confidence levels constrained by the Pantheon binned data only, and the orange contours show those with the inclusion of the VHZ SNe Ia. The gray and orange vertical dashed lines show the 1$\sigma$ confidence levels of the Pantheon binned data only and those of the combined data, respectively. The numbers of the image titles show the 1$\sigma$ confidence of the combined data. The contour levels are 1, 2, and 3-$\sigma$ with decreasing color weight.}

\label{fig:NoEvolw0waCDMCMB}

\end{figure}

The combination with the cosmological constraints of the Planck 2018 publication \citep{Planck182020A&A...641A...6P} is shown in Figure~\ref{fig:NoEvolw0waCDMCMB} for the Flat $w_0w_a$CDM. For the original Pantheon data and the VHZ SNe~Ia combined, the addition of the Planck 2018 measurements leads to  estimated values of $w_0 =-1.03^{+0.12}_{-0.11}$ and $w_a = 0.076^{+0.47}_{-0.55}$, which shows a significant improvement compared to the values without the Planck 2018 prior, as shown in Figure~\ref{fig:NoEvolw0waCDM}. They represent an optimistic expectation of the constraints on cosmological parameters that can be derived from the combined SNIa data based on the combination of the VHZ SNe~Ia with currently available observed data. The main comparison data points for the VHZ SNe~Ia are likely the BAO measurements being collected by the HETDEX \citep{HETDEX2008ASPC..399..115H} and DESI \citep{DESI2016arXiv161100036D} projects. It is important to have two independent measurements for such key parts of the cosmological history.

The extrapolations to increasingly larger samples of lower redshift SNe are shown in Figure~\ref{fig:NoEvolw0waCDMCMB} for the Flat $w_0w_a$CDM, where the statistical errors of the Pantheon data are reduced by an amount of $\sqrt{N}$, for $N$ =1, 4, 8,  and 64. The Figure shows that errors on $w_0$ and $w_a$ can be reduced by including the VHZ SNIa data. A realistic expectation in the future is $N$ =4 or 8, which will achieve the goals of the Stage IV SN experiments of the DETF \citep{DETF2006astro.ph..9591A} if systematic intrinsic evolution of SNIa luminosity distance can be neglected. 

\subsection{Models with Systematic Redshift Evolution of the Intrinsic Properties of SNe~Ia}

The discussions in \S~\ref{subsection:NoEvol} assume no redshift dependence of the distances determined from SNe~Ia. They assume that there are no other physical processes than the assumed cosmological models affecting the measured distances. 
The real power of the VHZ SNe~Ia resides in the control of systematic errors due to the evolution of supernova properties unaccounted for in the standardization processes. There are multiple ways the VHZ SNe~Ia can be used for such systematic studies. For the systematic evolution of SN properties, because the VHZ SNe are drawn from a population of SNe with the youngest progenitor age and lowest metallicity the universe has ever produced, they are likely either a subset of SNe~Ia at lower redshifts or have evolved smoothly through the cosmic expansion but with no counterparts at redshift zero. Several methods have been developed in recent years to identify objects based on their observed properties. The Twins Embedding of SNe~Ia  \citep{TwinsEmbeddingI2021ApJ...912...70B,TwinsEmbeddingII2021ApJ...912...71B} in particular, can be searched and used for cosmological inferences to minimize systematic effects arising from the evolution of SNe~Ia properties with redshifts. The AIAI technique allows for data-driven matching of observations and theories \citep{ChenXingzhuoMuse2021arXiv210106242C} which may enable physics informed determination of SNIa distances. The VHZ SNe are the most robust and straightforward sub-group of SNe that will allow decoupling of the age and metallicity effects when compared with lower redshift SNe. The methods for disentangling SNIa populations are not explored here. The broad redshift coverage on the Hubble diagram allows the VHZ SNe~Ia to constrain the systematic effect directly by parameterized likelihood fitting. 

We will explore two forms of the systematic effect, a linear relation to redshift given by $\delta \mu(z) = \hat{\Gamma}_1 z$ and a logarithmic relation given by $\delta \mu(z) = \hat{k}\ln(1+z)$. The parameters  $\hat{\Gamma}_1$
and $\hat{k}$ are determined simultaneously with the cosmological parameters, and for the fits involving the VHZ SNe~Ia, one more parameter $\hat{\Gamma}_0$ is applied to account for a potential zero point difference between the binned Pantheon data and the VHZ SNe~Ia. 

\begin{figure}
\includegraphics[width=0.45\textwidth]{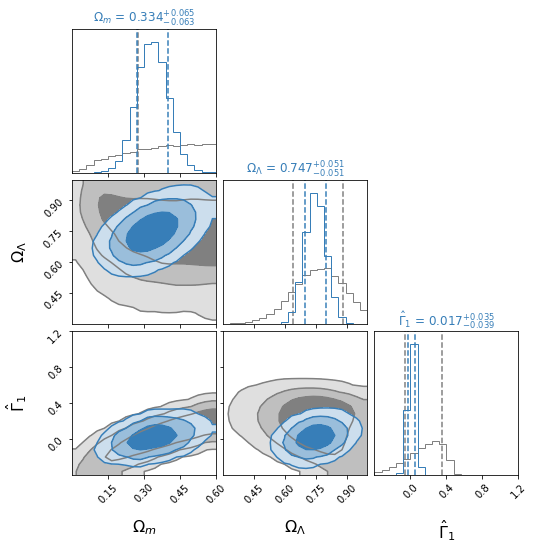}
\includegraphics[width=0.55\textwidth]{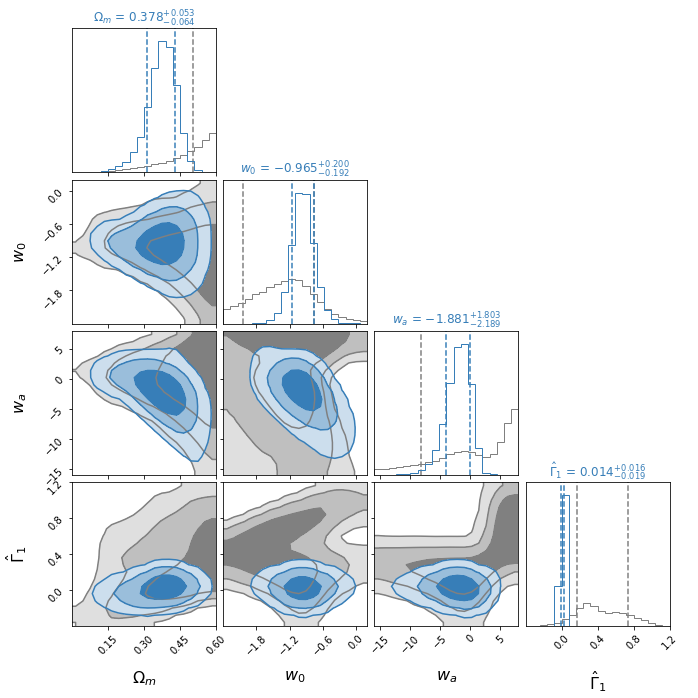}
\includegraphics[width=0.45\textwidth]{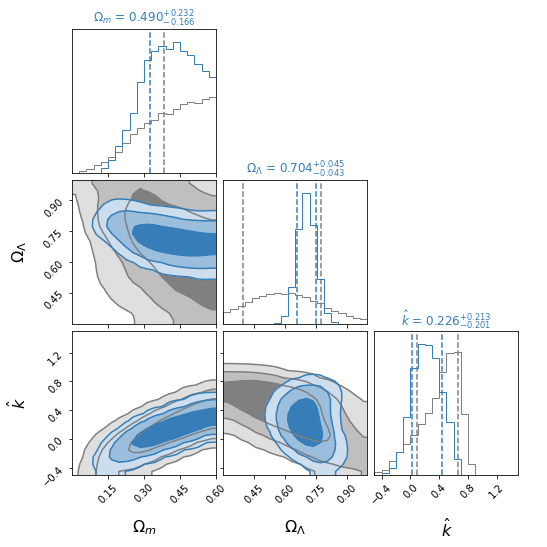}
\includegraphics[width=0.55\textwidth]{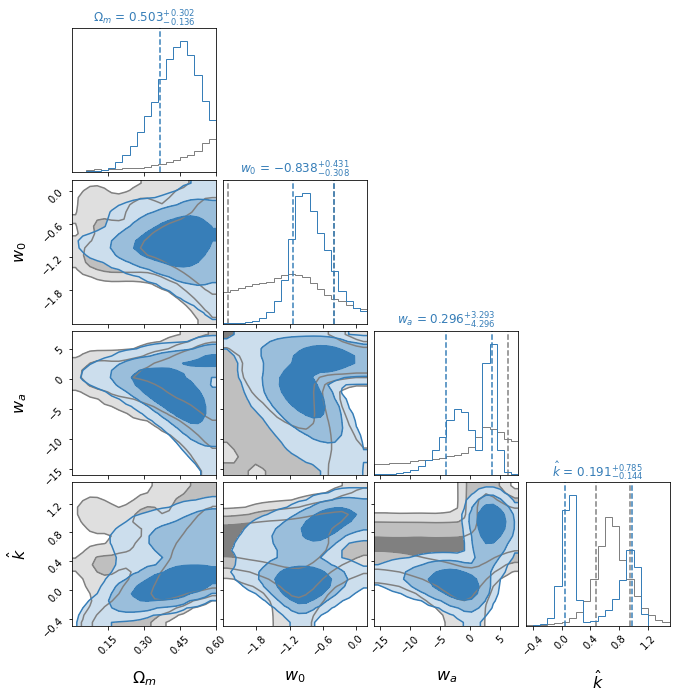}
\caption{The upper and lower two panels show the model fits assuming a linear ($\delta \mu(z) = \hat{\Gamma}_1 z$) and a logarithmic ($\delta \mu(z) = \hat{k}\ln(1+z)$) systematic evolution of the SNIa distances, respectively. The left and right panels are for the $\Lambda$CDM and Flat $w_0w_a$CDM models, respectively. The gray contours are for binned Pantheon data only, and the red for the joint data set including the VHZ SNe~Ia. The contours  with decreasing color weights are the 1, 2, and 3-$\sigma$ confidence regions.}
\label{fig:linlogEvol}
\end{figure}

The results for $\Lambda$CDM and Flat $w_0w_a$CDM are shown in Figure~\ref{fig:linlogEvol} for the case with no Planck prior. 
We found that the constraints on cosmological parameters deteriorate sharply when only the Pantheon data are used. This can be understood as most of the Pantheon SNe are at redshifts below 1; the assumed systematic effect is difficult to be disentangled from the cosmological models in such a small redshift range. The systematic evolution of SN distance moduli with redshift needs to be controlled from independent evidence to a negligibly small level for the lower redshift SNIa data to tightly constrain the cosmological parameters. The inclusion of the VHZ SNe~Ia restores the confidence levels of the cosmological parameters to the levels comparable to what can be achieved by the existing Pantheon data in the ideal case of no systematic redshift evolution of SNIa luminosity distances. We note also from the lower right panel of Figure~\ref{fig:linlogEvol} that there are very strong covariances between the parameter $\hat{k}$ for the assumed logarithmic evolution and the cosmological parameters $w_0$ and $w_a$ which makes the probability distribution of $w_a$ bimodal. This is likely caused by the errors in the assumed covariance matrix of the Pantheon supernova data. At the precision shown in Figure~\ref{fig:linlogEvol}, the results are extremely sensitive to the systematic error of the observational data.

\begin{figure}
\includegraphics[width=0.45\textwidth]{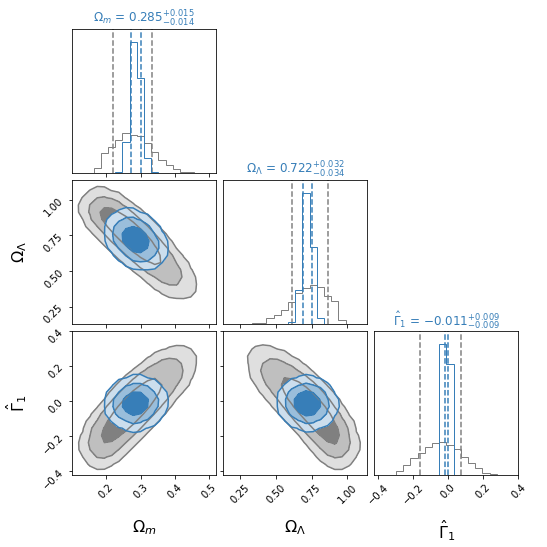}
\includegraphics[width=0.55\textwidth]{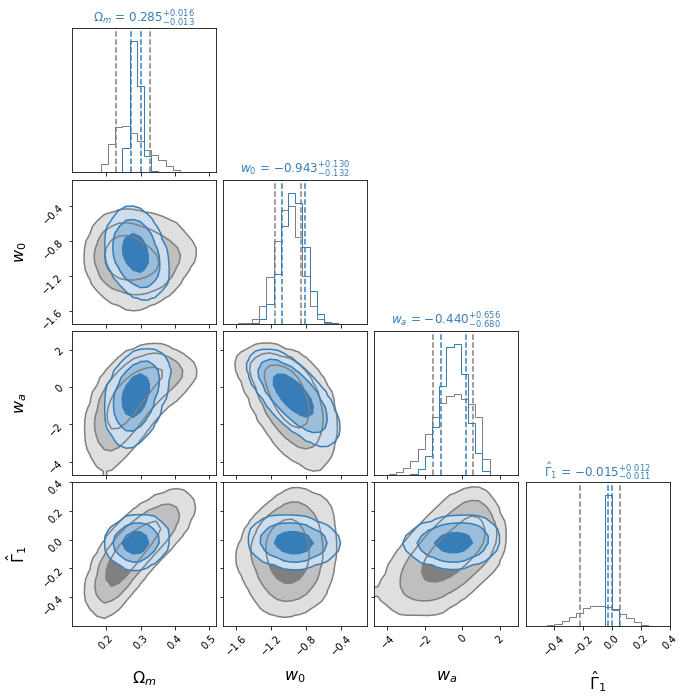}
\includegraphics[width=0.45\textwidth]{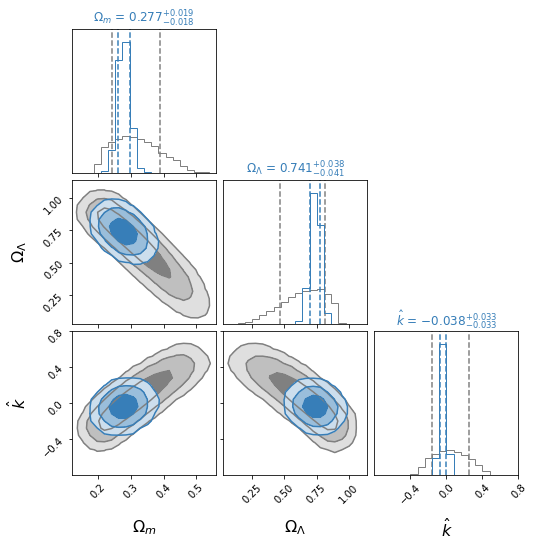}
\includegraphics[width=0.55\textwidth]{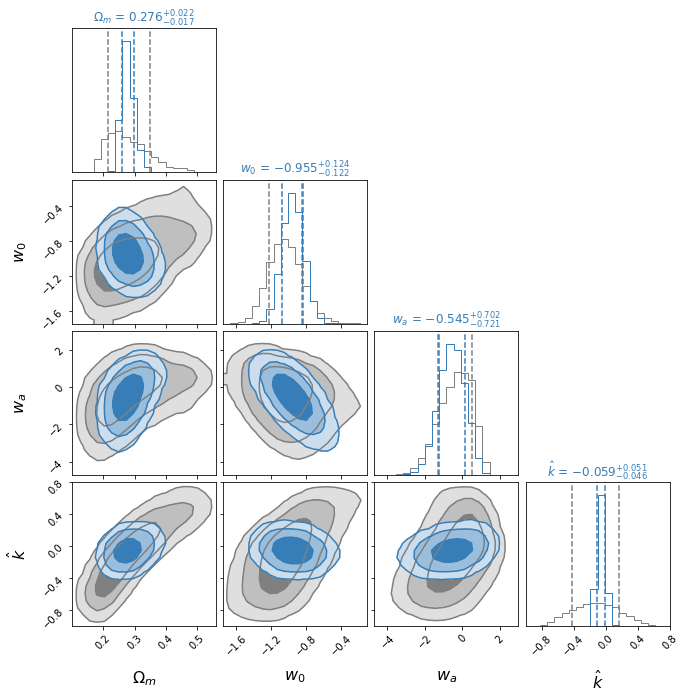}
\caption{Same as Figure~\ref{fig:linlogEvol}, but combined with the Planck 2018 results.
\label{fig:linlogEvolCMB}}
\end{figure}

Since both the CMB and the VHZ SNe~Ia provide tight constraints on $\Omega_M$, one may expect that the VHZ SNe may be replaceable by the Planck data. Indeed, the improvement to cosmological parameter constraints is only moderately observed after the inclusion of the VHZ SNe~Ia as shown in Figure~\ref{fig:NoEvolw0waCDMCMB}. This indicates that for cosmological parameters only, there is a considerable degeneracy of information provided by the VHZ SNIa data and the Planck data. 
However, the inclusion of a parameterized effect of redshift evolution other than cosmological models changes the conclusion completely. This is shown in Figures~\ref{fig:linlogEvol} and \ref{fig:linlogEvolCMB}, with details of the parameters shown in Tables~\ref{tab:paramsLCDM} and \ref{tab:paramsFlatw0waCDM}. 

Indeed, the combination of Pantheon and Planck 2018 data improves the parameter constraints, as shown by comparing the gray contours in Figure~\ref{fig:linlogEvol} for the $\Lambda$CDM and Flat $w_0w_a$CDM to those in Figure~\ref{fig:linlogEvolCMB}. But in all cases, the hypothetical evolutionary effect cannot be determined to a competitive level without the VHZ SNe~Ia. The inclusion of the VHZ SNe tightens the constraints dramatically, leading to constraints up to the levels of Stage IV supernova experiments \citep{DETF2006astro.ph..9591A}. The Planck data also effectively break the degeneracy (see Figure~\ref{fig:linlogEvolCMB}) of the bimodal probability distribution shown in Figure~\ref{fig:linlogEvol} for the Flat $w_0w_a$CDM model with a logarithmic SNIa distance modulus evolution. 

\begin{figure}
\includegraphics[width=0.45\textwidth]{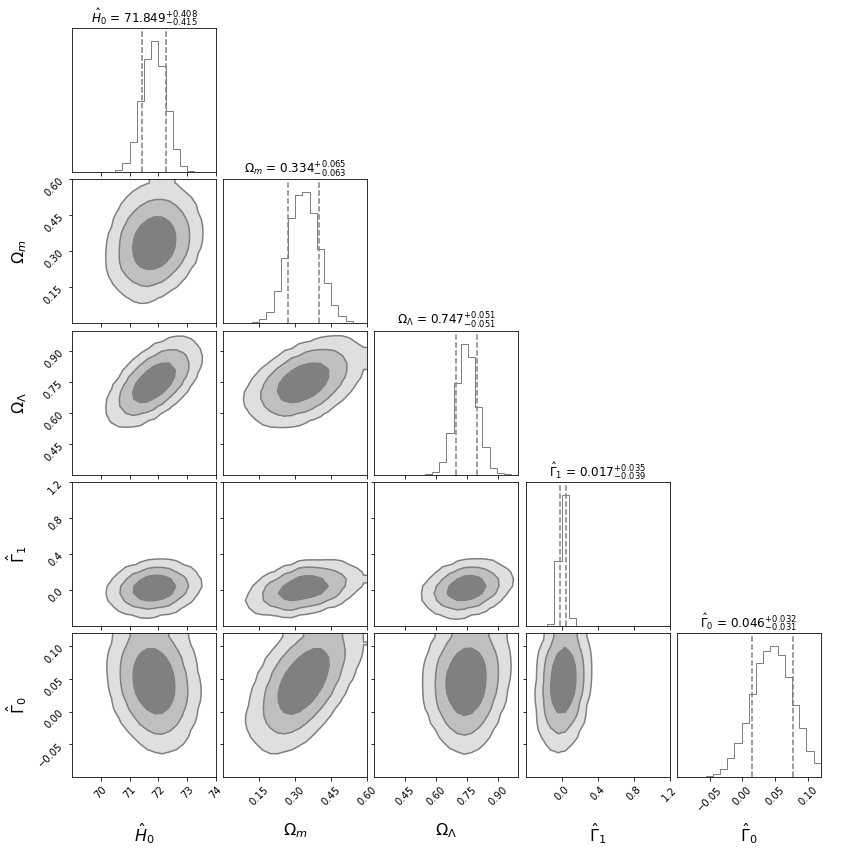}
\includegraphics[width=0.55\textwidth]{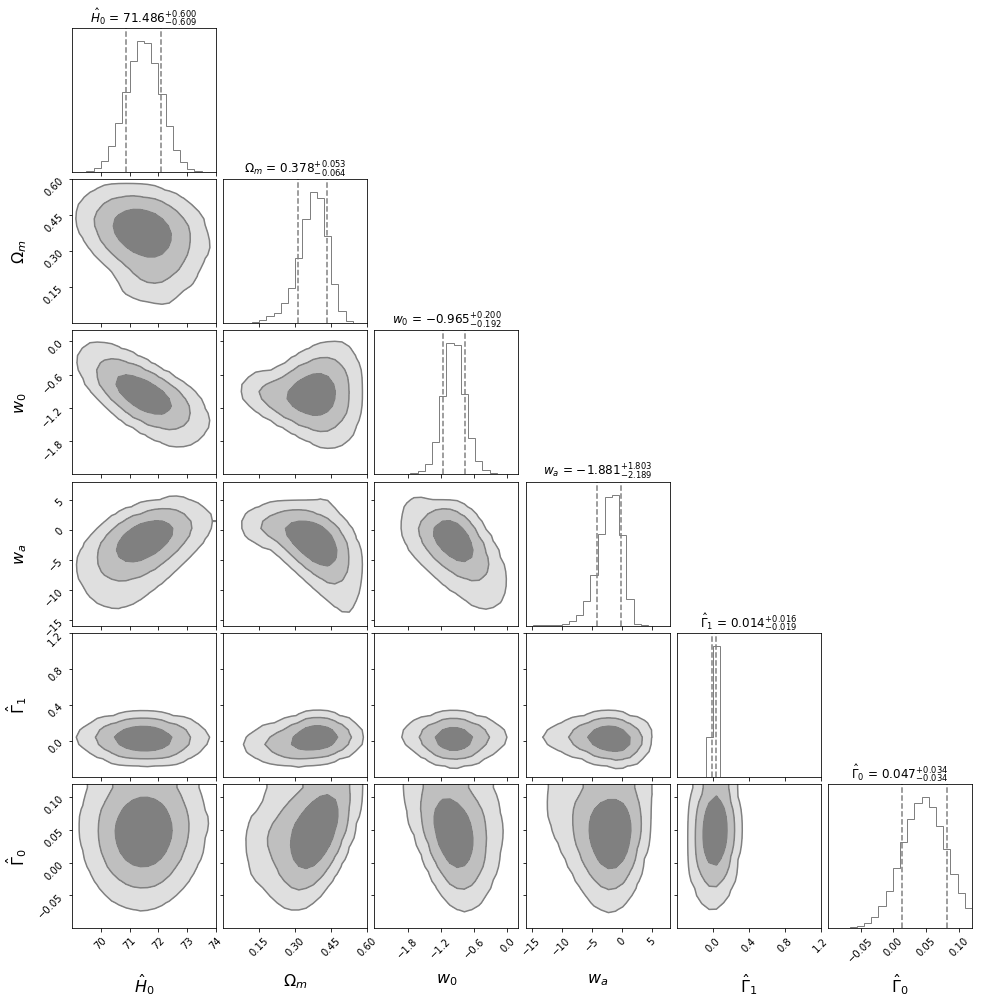}
\includegraphics[width=0.45\textwidth]{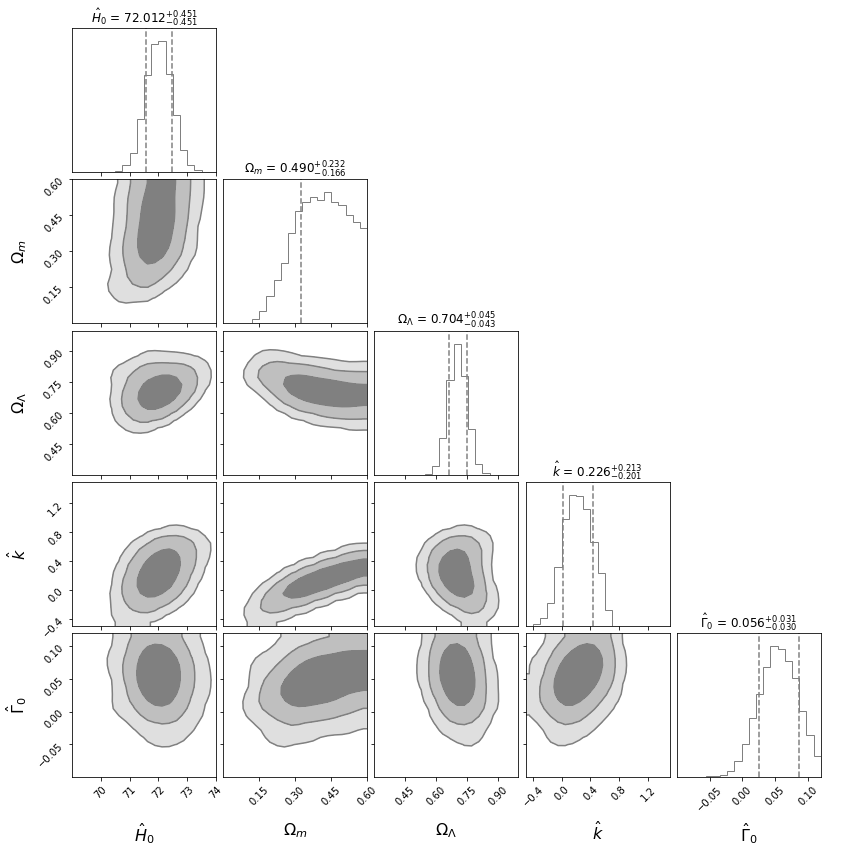}
\includegraphics[width=0.55\textwidth]{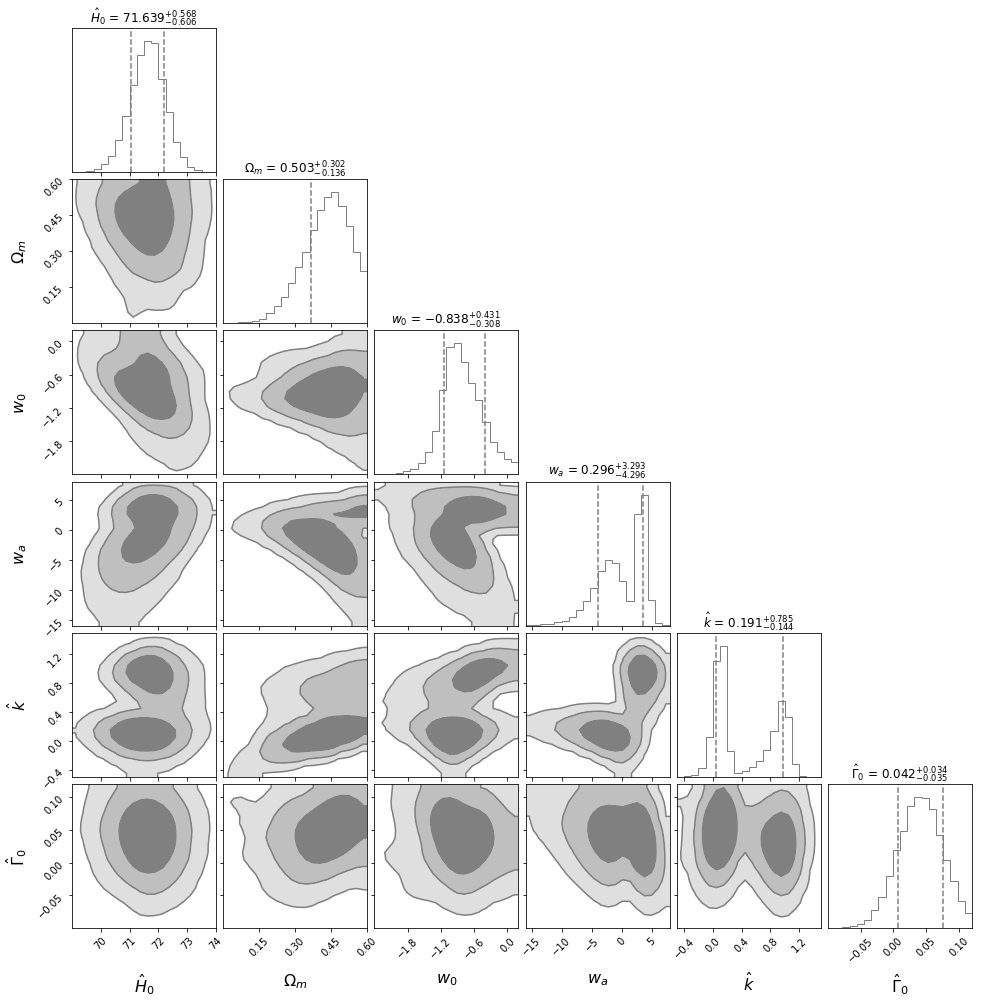}
\caption{The upper and lower two panels show the model fits, assuming a linear and a logarithmic systematic evolution of the SNIa distances, respectively. The left and right panels are for the $\Lambda$CDM and Flat$w_0w_a$CDM models, respectively. The contours are calculated for the combined binned Pantheon data and the VHZ SNe Ia. The titles of the corner plot boxes show the 1 $\sigma$ range of the parameters. The contour levels are 1, 2, and 3-$\sigma$ with decreasing color weight.}
\label{fig:linlogEvol_A}
\end{figure}

It is also interesting to know how well the additional nuisance parameters introduced to model the systematic evolution can be constrained. Figure~\ref{fig:linlogEvol_A} shows the confidence levels of all the parameters involved in the fits for the $\Lambda$CDM and Flat $w_0w_a$CDM models. First, we see that the bias level $\hat{\Gamma}_0$ between the VHZ SNe~Ia and the Pantheon data can be determined to better than 0.035 mag for all these models. This implies that no significant deterioration of the cosmological fits is expected if the photometric system of the VHZ SNIa data set is calibrated to within 0.035 mag of the Pantheon data. The coefficient $\hat{\Gamma}_1$ can be determined to $\sim$ 0.04, which implies a systematic shift of $\delta m \sim 0.20$ mag at $z \sim$ 5. This value is close to typical Hubble residuals of SNIa luminosity distances. The value $\hat{k}$ can be determined to a precision of $\sim$ 0.22, which translates to a $\delta m\ \sim 0.39$ mag at $z \sim$ 5 and is larger than typical values of the observed intrinsic dispersion of the standardized SNIa luminosities. This suggests that the models for the logarithmic evolution may be further improved if a tighter prior on the systematic evolution is set based on existing SNIa data.

\begin{figure}

\includegraphics[width=0.45\textwidth]{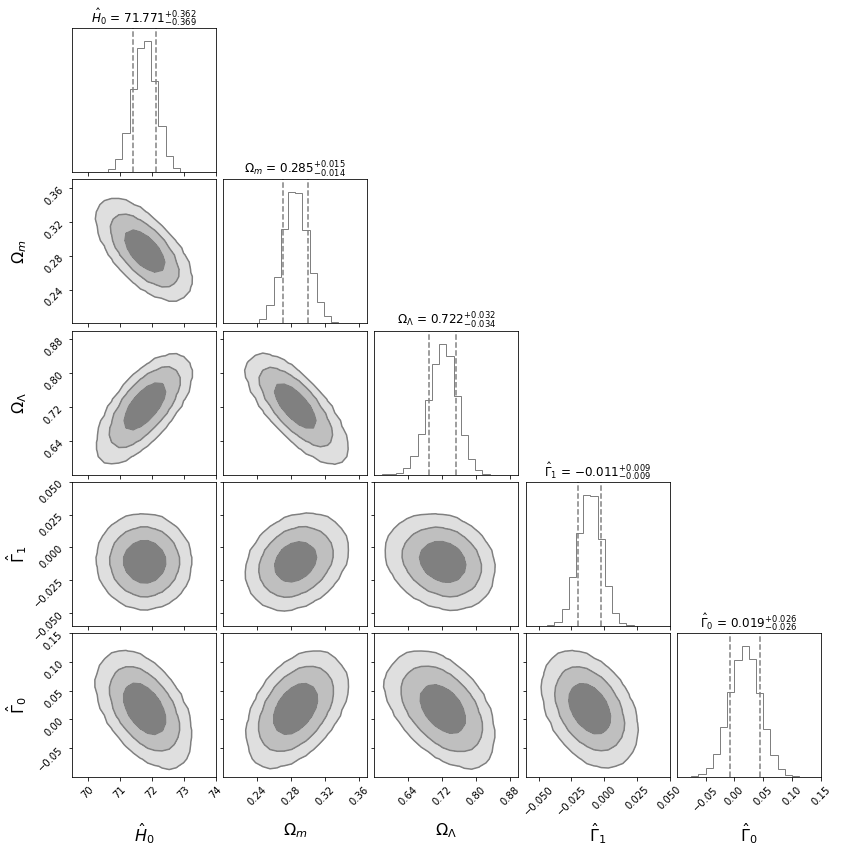}
\includegraphics[width=0.55\textwidth]{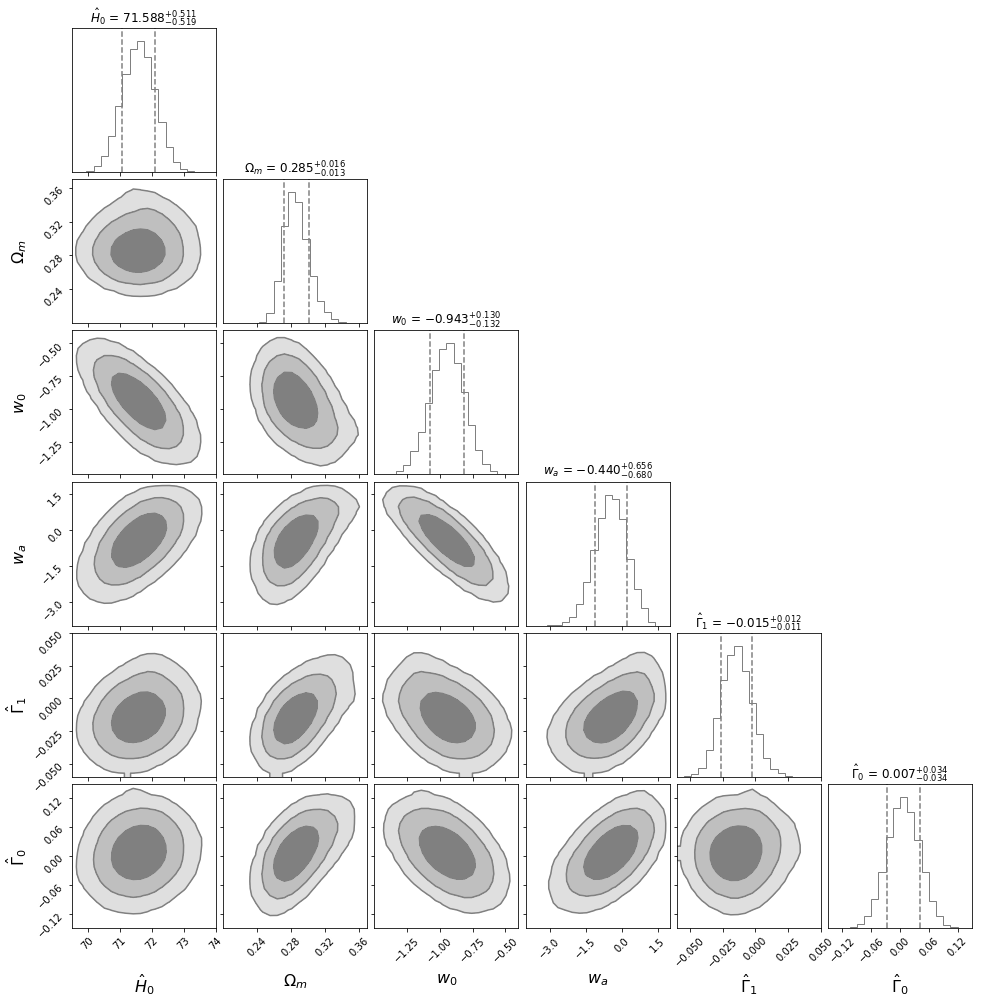}
\includegraphics[width=0.45\textwidth]{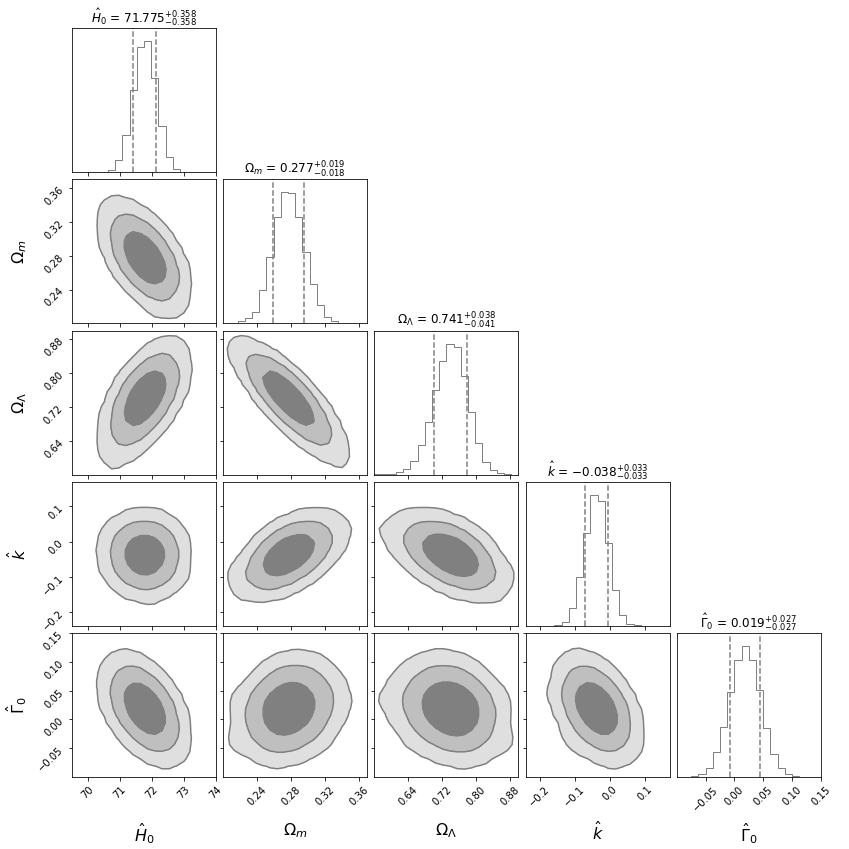}
\includegraphics[width=0.55\textwidth]{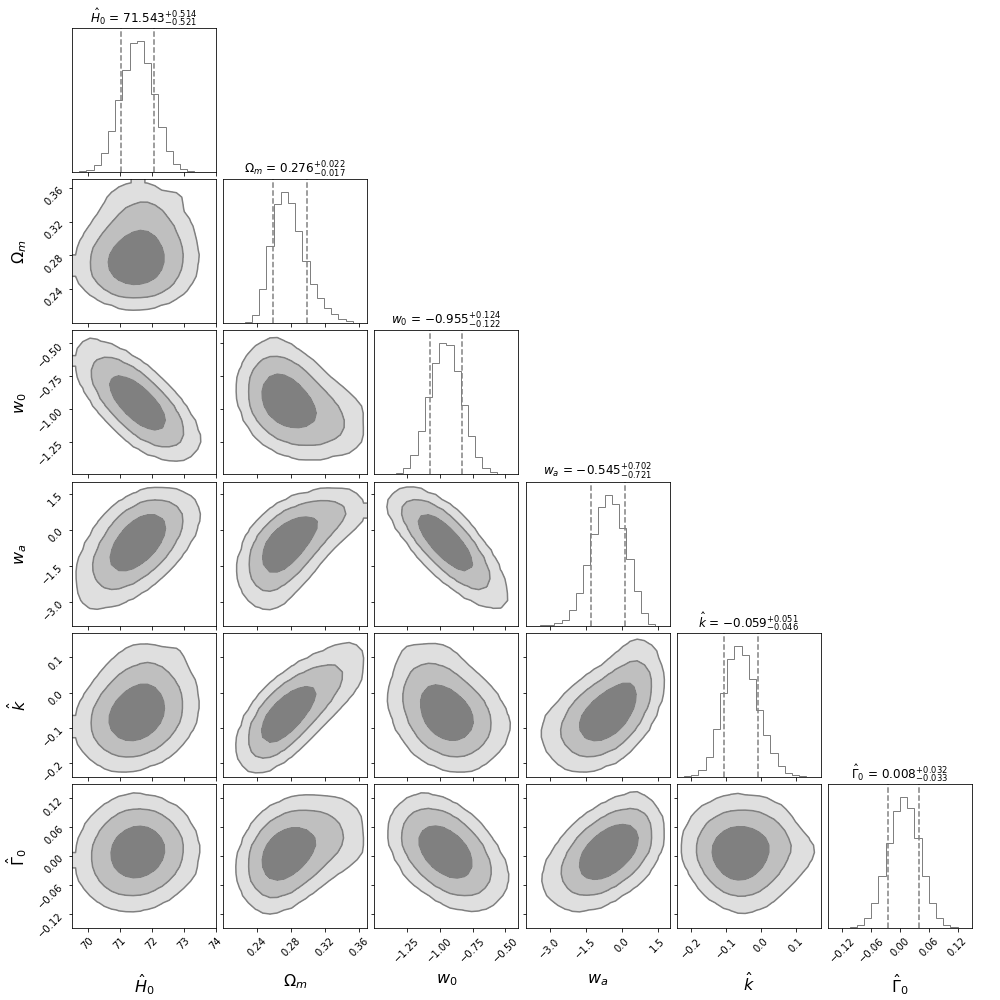}

\caption{Same as in Figure~\ref{fig:linlogEvol_A} but combined with the Planck 2018 results.
\label{fig:linlogEvol_ACMB}}
\end{figure}

The combination with Planck measurements tightens these nuisance parameters further (see Figure~\ref{fig:linlogEvol_ACMB}): The parameters $\hat{\Gamma}_1$ and $\hat{k}$ can be determined to levels of $\pm 0.009$ and $\pm 0.033$ for the $\Lambda$CDM, which implies that a systematic evolution of the supernova distance scale will be disentangled from the cosmological effect down to levels of 0.01 mag and 0.023 mag at $z \sim$ 1 for the linear and logarithmic evolution relations, respectively. These values are well below typical magnitude residuals of Hubble diagrams constructed using SNIa data at $z\sim 1$ and most known intrinsic scatters of SNIa distance standardization. In contrast, the effect of systematic evolution is indistinguishable from the cosmological effects of the $\Lambda$CDM model at 0.14 mag and 0.19 mag levels for the linear and logarithmic evolution with the Planck data but without the VHZ SNe~Ia, respectively (see Table~\ref{tab:paramsLCDM}). Similar effects can be seen for other cosmological models, as shown in Tables~\ref{tab:paramsFlatw0waCDM} and \ref{tab:paramsFlatw0waCDMX16}. The VHZ SNe~Ia will thus lay a solid foundation for precision cosmology by providing tight control of the systematic redshift evolution unrelated to cosmology.

\begin{figure}
\includegraphics[width=0.45\textwidth]{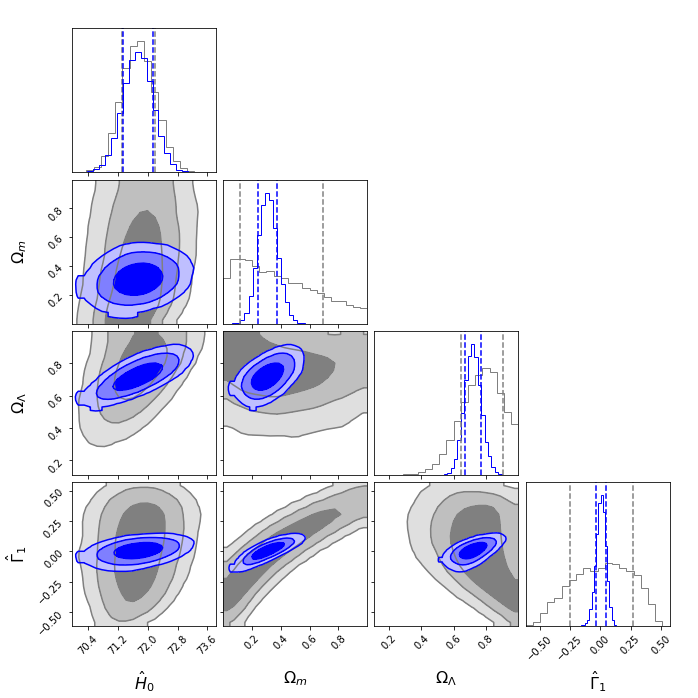}
\includegraphics[width=0.45\textwidth]{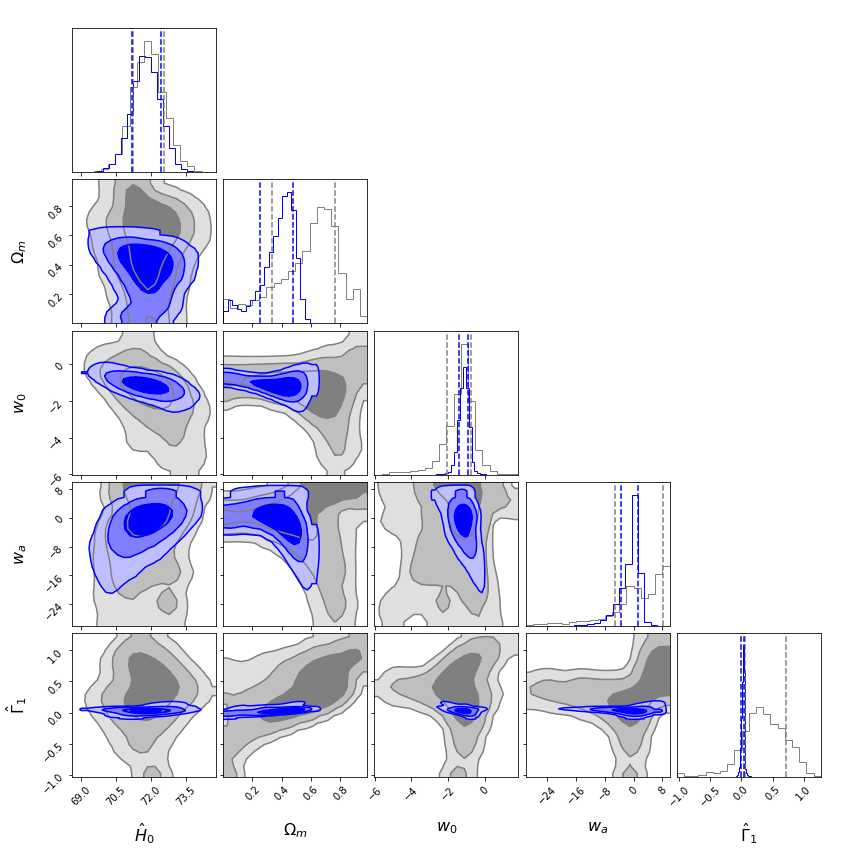}

\caption{The parameter constraints for the $\Lambda$CDM and Flat $w_0w_a$CDM assuming a linear systematic evolution ($\delta\mu=\hat{\Gamma}_1z$) of the SN distance modulus with redshifts for the Pantheon only data (gray) and the joint data including the VHZ SNe (blue). The vertical dashed lines show the 1 $\sigma$ confidence levels. The contour levels with decreasing color weights are the 1, 2, and 3 $\sigma$ confidence regions.}
\label{fig:EvoquadlogBlue}
\end{figure}

For comparison, the probability distributions of cosmological parameters are presented again in Figure~\ref{fig:EvoquadlogBlue} 
for the $\Lambda$CDM and Flat $w_0w_a$CDM, the contours with the combined Pantheon and VHZ SNIa data are shown (in blue) together with those that can be derived for the Pantheon data only (in gray). Note that for the combined data, an additional fitting parameter $\hat{\Gamma}_0$ is included but not plotted.  Even with two additional fitting parameters, adding VHZ SNe~Ia leads to significantly improved constraints on the cosmological parameters. The VHZ SNe~Ia anchor $\Omega_M$ to a value with a much higher precision which improves the constraints on dark energy related parameters such as $\Omega_\Lambda$, $w_0$ and $w_a$. 
The Pantheon data alone do not allow for meaningful constraints on $\hat{\Gamma}_1$ or $\hat{k}$ that describe the systematic evolution of the SN magnitudes with redshifts. 
The combined data set, however, is sensitive to $\hat{\Gamma}_1$ values to  $0.014^{+0.016}_{-0.019}$ for the Flat $w_0w_a$CDM model in Table~\ref{tab:paramsFlatw0waCDM}. Since the central value of $\hat{\Gamma}_1$ is 0 by construction, only the derived error levels are important. The fits shown in Figures~\ref{fig:linlogEvol} and \ref{fig:EvoquadlogBlue} suggest that systematic evolution with $z$ larger than $|\sim 0.019z|$ can be self-calibrated and separated from other cosmology related parameters. At $z \sim 5$, this translates to a magnitude evolution of $0.095$ mag. In the extreme case that the first generation SNe~Ia are the dimmest or brightest extremes of their lower redshift counterparts, the VHZ SNe~Ia would enable their effects on cosmological parameter determination to be quantified and eliminated.

Note that in Tables~\ref{tab:paramsLCDM} and \ref{tab:paramsFlatw0waCDM} the constraints on the bias level $\hat{\Gamma}_0$ have a value consistent with 0 but with errors $\sim$ 0.035.  The SNe were generated using the results of a Flat $\Lambda$CDM fit to the Pantheon data. The $\hat{\Gamma}_0$ derived from the model fit is thus consistent with the initial setup. The small error values are interesting as they suggest that if for some unknown reasons the distances of the VHZ SNe~Ia have a systematic offset compared to the CLZ SNe~Ia, we would be able to detect that offset to within a level of 0.035 mag at a  1 $\sigma$ confidence level. It also implies that it will not severely impact the precision of the cosmological inferences if $\hat{\Gamma}_0$ is smaller than 0.035 mag. This provides us with a calibration goal for the VHZ SNe~Ia.

The contours of Figures~\ref{fig:linlogEvol_A}, \ref{fig:linlogEvol_ACMB} and \ref{fig:EvoquadlogBlue} show strong $\hat{\Gamma}_1$--$\Omega_M$ and $\hat{k}$--$\Omega_M$ covariances. This is consistent with the nearby SNIa data being unable to constrain $\hat{\Gamma}_1$ and $\hat{k}$ effectively. The VHZ SNIa sample breaks such covariances and sets stronger constraints on both $\Omega_M$ and the evolutionary parameters. 

From Table~\ref{tab:paramsLCDM} we see that in the ideal case where the systematic evolution of SNIa luminosity distances is well under control during the standardization process, the VHZ SNe~Ia provide nearly identical constraining power as can be provided by the Planck data. The $\Omega_M$ and $\Omega_\Lambda$ errors from Pantheon with Planck Prior, Pantheon and VHZ SNe~Ia, and Pantheon and VHZ SNe~Ia with Planck Prior are nearly identical. If a fit to the systematic evolution is needed, the Pantheon and VHZ SNe~Ia lead to significantly improved results. In particular, the Pantheon and VHZ SNe~Ia with Planck prior yield errors that are identical without the addition of the linear evolution term, suggesting that a linear redshift evolution can be self-calibrated by statistical analysis of the data.  In Table~\ref{tab:paramsFlatw0waCDM}, we can see that similar behavior, a potential systematic redshift evolution of the SNIa luminosity distance, even at a very low level, may significantly weaken cosmological constraints based only on SNIa in the Pantheon redshift range. With $\hat{\Gamma}_1$ being $-0.055^{+0.039}_{-0.040}$ as in the case of Pantheon with Planck prior, the uncertainty of the magnitude of the evolutionary effect is $\sim 0.04$ at $z=1$ which is close to the level of the magnitude dispersion of the Hubble residuals of typical SNIa Hubble diagrams. 
The VHZ SNe~Ia can be instrumental in calibrating such systematic evolutionary effect and bring down the errors to around $\sim 0.02z$ and $\sim 0.01z$ without and with Planck prior, respectively. This allows for the extraction of cosmological parameters with the evolutionary effect reduced to a negligible level using the CLZ SNIa. Similar improvements can be seen also for the case of the logarithmic redshift dependence.

\begin{table}[]
\caption{Constraints on Cosmological Parameters of  Flat $w_0w_a$CDM with the Statistical Errors of Pantheon Data Reduced by a Factor of 4}
\begin{tabular}{l|cccccc}
\hline
\hline 
Data set  & $\Omega_M$ & $w_0$ &$w_a$ & $\hat{\Gamma}_0$ & $\hat{\Gamma}_1$ & $\hat{k}$\\
\hline
Pan16\tablenotemark{a}   & $0.397_{-0.075}^{+0.050}$ & $-0.954_{-0.162}^{+0.149}$  & $-1.920_{-2.203}^{+1.742}$ &  \\
Pan16+P\tablenotemark{a}   & $0.283_{-0.003}^{+0.003}$ & $-0.778_{-0.060}^{+0.060}$&
$-1.124_{-0.254}^{+0.241}$ & \\
Pan16VHZ\tablenotemark{a}  & $0.333_{-0.051}^{+0.032}$ & $-0.953_{-0.103}^{+0.115}$&
$-0.530_{-1.191}^{+1.184}$ & $0.037_{-0.029}^{+0.028}$ & \\
Pan16VHZ+P\tablenotemark{a}  & $0.282_{-0.004}^{+0.004}$ & $-0.799_{-0.061}^{+0.061}$&
$-1.061_{-0.259}^{+0.246}$&
$-0.053_{-0.011}^{+0.011}$ & \\
\hline
Pan16\tablenotemark{b} &  $0.708_{-0.247}^{+0.086}$ & $-1.311_{-0.591}^{+0.447}$  & $7.503_{-9.531}^{+1.840}$ & & $0.561_{-0.467}^{+0.156}$& \\
Pan16+P\tablenotemark{b} &  $0.282_{-0.004}^{+0.004}$ & $-0.947_{-0.105}^{+0.106}$  & $-0.486_{-0.407}^{+0.387}$ & & $-0.060_{-0.030}^{+0.030}$& \\
Pan16VHZ\tablenotemark{b}& $0.427_{-0.057}^{+0.048}$ & $-0.970_{-0.164}^{+0.166}$  & $-2.320_{-2.091}^{+1.674}$ &  $0.049_{-0.030}^{+0.030}$ &  $0.030_{-0.017}^{+0.014}$\\
Pan16VHZ+P\tablenotemark{b} & $0.282_{-0.003}^{+0.004}$ & $-0.833_{-0.061}^{+0.063}$  & $-0.910_{-0.260}^{+0.244}$ & $-0.012_{-0.022}^{+0.022}$ & $-0.017_{-0.008}^{+0.008}$\\
\hline
Pan16\tablenotemark{c}&  $0.836_{-0.167}^{+0.088}$ & $-1.046_{-0.850}^{+0.970}$  & $4.438_{-7.171}^{+3.831}$   & & & $0.848_{-0.304}^{+0.172}$ \\
Pan16+P\tablenotemark{c}&  
$0.282_{-0.003}^{+0.004}$ & $-0.967_{-0.125}^{+0.120}$  & $-0.403_{-0.462}^{+0.457}$  & & & $-0.085_{-0.050}^{+0.046}$ \\
Pan16VHZ\tablenotemark{c} & $0.507_{-0.092}^{+0.077}$ & $-0.947_{-0.213}^{+0.218}$  & $-2.887_{-3.180}^{+2.412}$  &$0.046_{-0.030}^{+0.029}$ & & $0.161_{-0.078}^{+0.068}$ \\
Pan16VHZ+P\tablenotemark{c}&  $0.282_{-0.003}^{+0.004}$ & 
$-0.913_{-0.073}^{+0.073}$  & $-0.600_{-0.290}^{+0.277}$  & $0.014_{-0.027}^{+0.027}$ & & $-0.060_{-0.022}^{+0.022}$ \\
\hline
\end{tabular}
\begin{itemize}
\small{
 \item[] \tablenotetext{}{Notes to Table entries: Pan16 stands for Pantheon data with statistical error reduced by a factor of 4, Pan16+P for Pan16 with Planck prior, Pan16VHZ for Pan16 and the VHZ SNe~Ia, and Pan16VHZ+P for Pan16 and VHZ SNe~Ia with Planck prior. }
\item[] \tablenotetext{a}{No systematic evolution.}
\item[] \tablenotetext{b}{Linear systematic evolution proportional to $\hat{\Gamma}_1$z.}
\item[] \tablenotetext{c}{Logarithmic systematic evolution proportional to $\hat{k}\ln(1+z)$.}}
\end{itemize}
\label{tab:paramsFlatw0waCDMX16}
\end{table}

For a future perspective, one may expect the sample of local SNe~Ia to grow in size and quality. Table~\ref{tab:paramsFlatw0waCDMX16} shows the results with the statistical errors of the Pantheon sample reduced by a factor of 4 while keeping all the systematic errors unchanged, or equivalently, with the size of the SN sample increased by a factor of 16 but all sources of systematic errors controlled to the same level of the existing Pantheon data set. Such a data set with more stringent systematic error controls may be expected from future observations with the Rubin/LSST and Roman/WFIRST observatories. The central values of most parameters in Table~\ref{tab:paramsFlatw0waCDMX16} show large deviations from the assumed cosmology used to construct the mock VHZ SNIa data. This indicates that the fits are dominated by systematic errors of the lower redshift SNe~Ia. However, it is interesting to note that when all the SN data are combined with the Planck data, the cosmological parameters restore to values close to those assumed for the mock VHZ SNe~Ia. The errors on cosmological parameters from the joint SNIa and Planck fits with a linear or logarithmic luminosity distance evolution can again approach the levels ignoring such evolutionary effects.  

\section{Conclusions}
\label{sec:conclusion}

\begin{figure}

\includegraphics[width=0.5\textwidth]{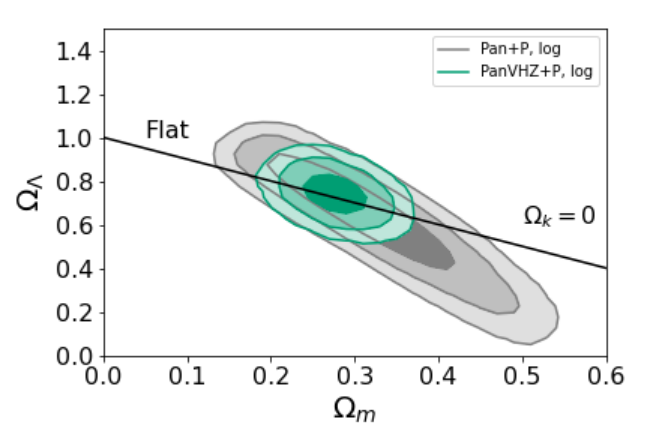}
\includegraphics[width=0.5\textwidth]{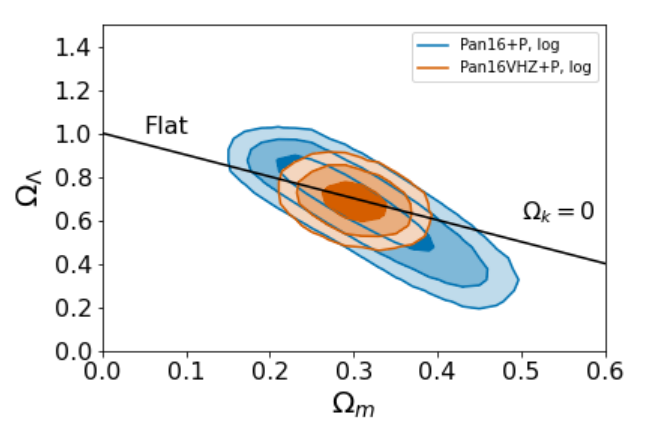}
\includegraphics[width=0.5\textwidth]{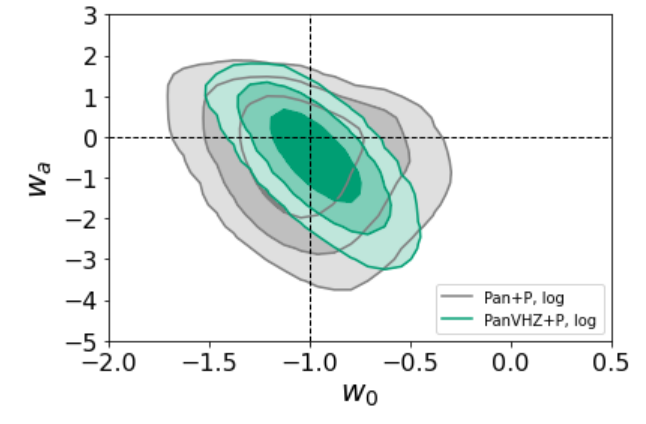}
\includegraphics[width=0.5\textwidth]{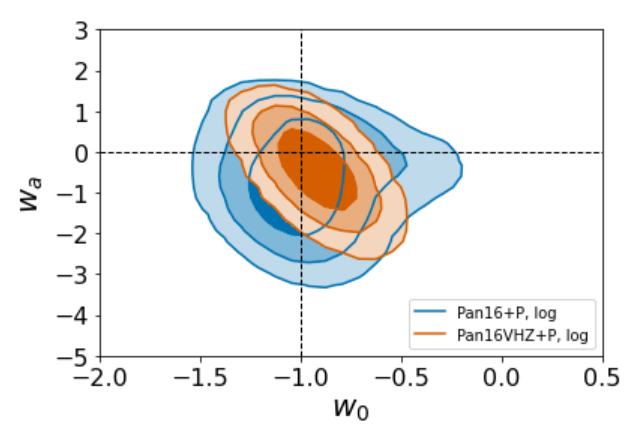}

\caption{Comparisons of the 
constraints to the $\Lambda$CDM (upper) and Flat $w_0w_a$CDM (lower) models assuming a logarithmic evolution ($\delta\mu=\hat{k}\ln(1+z)$) of the distance modulus with redshift. The contours are 1-$\sigma$, 2-$\sigma$ and 3-$\sigma$ levels for the Pantheon with Planck prior (gray lines), Pantheon plus VHZ SNe and Planck prior (green lines), Pantheon with statistical errors reduced by a factor of 4 and Planck prior (blue lines), and Pantheon with statistical errors reduced by a factor of 4 plus VHZ SNe (orange lines). The solid straight line on the upper panel shows the line for a flat universe. The solid dashed horizontal and vertical lines on the lower panel show the assumed values of the $w_0$ and $w_a$ for the simulated VHZ SN data.}

\label{fig:summaryfig2}
\end{figure}

We have carried out MCMC simulations of cosmological fits using an SNIa data set consisting of the existing Pantheon compilation and a mock VHZ SNIa sample that can be obtained by the JWST in the coming years. We examined the cases without and with the assumption of an explicit form of systematic evolution. We found that for the simple cases of a linear and a logarithmic SN luminosity distance redshift evolution, a joint cosmological fit will successfully eliminate the evolutionary effects with the addition of the VHZ SNe~Ia.

The major results are summarized in Figure~\ref{fig:summaryfig2} using the $\Lambda$CDM and Flat $w_0w_a$CDM with a logarithmic SN magnitude evolution as examples. It is seen that even if the size of the Pantheon sample is enlarged by a factor of 16, it would still not be able to constrain the systematic evolution of SN magnitudes and the cosmological parameters as effectively as the VHZ SN data. Such a large size is only expected in future surveys with the LSST/Rubin Observatory.

Being able to establish the evolutionary effect will lead to significant improvements in cosmological parameter measurements. Such improvement can be extremely important for SN cosmology in the upcoming years with Rubin/LSST and Roman/WFIRST observatory, which will produce a significantly larger data set of well-observed SNe~Ia out to z $<$ 1.7. The statistical power of such a data set will need to be confronted with a detailed analysis of systematic effects. The VHZ SNe~Ia provide a unique capability to underpin the evolutionary effect of SNe~Ia.

\acknowledgments
We thank Eric Linder for helpful discussions. LW acknowledges support from an NSF grant AST-1817099. The work of SP was supported in part by the Director, Office of
Science, Office of High Energy Physics of the US Department
of Energy under contract No. DE-AC025CH11231. 
JV and ER were supported by the project GINOP-2.3.2-15-2016-00033 of the National Research, Development and Innovation Office (NKFIH), Hungary, funded by the European Union.




\software{Astropy\citep{2013A&A...558A..33A}, Corner\citep{2016JOSS....1...24F}, emcee\citep{emcee2013PASP..125..306F}, Matplotlib\citep{Hunter:2007}, MC3\citep{2017AJ....153....3C}, Numpy\citep{harris2020array}, Pandas\citep{mckinney-proc-scipy-2010}, Scipy\citep{2020NatMe..17..261V}}



\bibliography{VHZSNe}{}
\bibliographystyle{aasjournal}

\end{document}